\documentclass{article}

\PassOptionsToPackage{numbers,compress}{natbib}
\usepackage[preprint]{neurips_2026}

\usepackage[english]{babel}
\usepackage[utf8]{inputenc}
\usepackage[T1]{fontenc}
\usepackage{url}

\usepackage{amsmath}
\usepackage{tabularx}
\usepackage{booktabs}
\usepackage{multirow}
\usepackage{xcolor}

\usepackage{graphicx}
\usepackage{subcaption}
\usepackage{float}
\usepackage{placeins}

\usepackage[colorlinks=true, linkcolor=blue, citecolor=blue]{hyperref}

\usepackage{caption}
\usepackage{algorithm}
\usepackage{algpseudocode}

\renewcommand{\arraystretch}{1.1}
\title{BlossomPsy: A User-Centric AI System for Adaptive and Engaging MBTI Personality Assessments}
\author{%
  Bingjia Huang\\
  University of Electronic Science and Technology of China
}

\begin{document}
\maketitle

\begin{abstract}
There has been growing public interest in understanding personality traits and emotional characteristics, as such knowledge helps individuals better accept themselves and manage negative emotions. While professional personality scales remain the standard tool for assessment, they are often perceived as tedious or inaccessible to the general public. AI-driven systems can make assessments more accessible, but it is
difficult to balance user engagement with predictive consistency in existing works. We tackle this challenge by introducing BlossomPsy, a user-friendly AI-driven MBTI assessment system. MBTI, a widely recognized but psychometrically debated personality framework, serves as the foundation for many recent systems. BlossomPsy integrates multi-turn dialogue and photo-based questions to enhance user engagement while supporting confidence-aware predictions. By combining deep learning, multi-armed bandit algorithms, and control theory, the system dynamically adapts to users' responses. In particular, photo-based questions are designed to increase interactivity and provide additional user information, thereby improving prediction confidence. Experiments involving both human volunteers and large language models (LLMs) provide preliminary evidence that BlossomPsy can produce stable predictions, with higher reported user satisfaction compared to MBTI-M (Chinese version), while maintaining comparable consistency with the reference scale.
\end{abstract}
\section{Introduction}
Over the years, public awareness of mental health has steadily increased \cite{olawade2024enhancing}.  By knowing themselves better, people are more able to recognize and handle emotional challenges. Yet, most psychological assessment tools still use formats that have remained largely unchanged for decades. Typically, individuals need to complete a set of choice questions, after which the system automatically calculates their scores \cite{hogan1991personality,bohane2017resilients,hough2008personality}. By comparing their scores with standardized manuals, the system can provide a structured result to people. However, this kind of assessment often suffers from limited user engagement, repetitive formats, and difficulty adapting to individual differences. Many users consider such assessments tedious, leading to low participation especially among young people \cite{merry2012effectiveness}.
To make the assessment engaging, psychologists have explored various interactive or visual-based methods for assessments over the decades. \citet{buck1966house} put forward the House-Tree-Person Test (HTP). In the test, individuals draw a house, a tree and a person; the drawings are then analyzed to gain insights into their personality, emotions and inner experiences. \citet{greenwald1998measuring} introduced Implicit Association Test (IAT), in which the strength of automatic associations between concepts is measured to assess potential bias or attitude.  Further, game-based assessments like \cite{song2020validation} are used to assess children's cognitive control.

Recent advances in artificial intelligence and human-computer interaction have created new opportunities to reimagine psychological testing. Some studies have used NLP techniques to support psychology evaluation, like empathy detection \cite{sharma2020computationalapproachunderstandingempathy}, mental illness and agony evaluation \cite{saha2022shoulder} and text-based personality assessment \cite{peters2024large,maharjan2025psychometric}. \citet{teferra2024screening,zhang2022natural} provide a general review of NLP-based psychology evaluation. Although these studies have opened up a new avenue for exploration, they often lack practical applicability \cite{laricheva2024scoping}. LLMs may further improve applicability.  Some researchers have aimed to improve the capabilities of large language models (LLMs) so that they can provide psychological assessment traditionally handled by psychologists \cite{jin2025applications,sartori2023language}. Related research includes psychological disorder detection \cite{chen2023empowering, shin2024using}, mental health counseling \cite{zhang2024cpsycoun,hu2025beyond}, personality assessment \cite{peters2024large,maharjan2025psychometric} and so forth. In addition, researchers have begun to focus on improving the AI-based psychological evaluations in a more creative way. Some studies have combined psychological measurement with interactive fiction games \cite{yang2024psychogat,zhang2024cpsycoun}. \citet{li2025psydipersonalizedprogressivelyindepth} developed PsyDI, a psychology chatbot that utilizes multi-turn dialogues to deliver customized assessments within the Myers-Briggs Type Indicator (MBTI) framework.
Despite these innovations, current systems still struggle to balance predictive consistency and user engagement. In principle, the reliability and validity of a newly designed scale should be examined through rigorous experimentation and data analysis \cite{devellis2021scale}. This process requires a solid foundation in psychometric scale development as well as a substantial sample size. However, due to constraints in time, financial resources and effort, many studies omit such testing, which weakens the empirical basis of their designs\cite{almakinah2025enhancing,wang2025generativeai}.

To address these limitations, we propose \textbf{BlossomPsy}, a hybrid multi-modal MBTI assessment system that combines traditional questionnaire structure with intelligent multi-turn conversation and adaptive visual prompts. BlossomPsy is intended as a research prototype for interactive personality assessment rather than a clinical diagnostic or mental-health treatment tool. Based on the MBTI framework, we incorporate a two-stage prediction strategy: conversations between users and LLMs are evaluated by a Multi-Head Classifier (MHC)---a shared-encoder architecture with multiple parallel classification heads---and an adaptive fallback module that presents \textbf{photo-based single-choice questions} when the model lacks sufficient confidence. Inspired by the multi-armed bandit (MAB) framework \cite{auer2002finite}, we design a modified Upper Confidence Bound (mUCB) algorithm to make MBTI predictions and guide the focus of the conversation, and a Proportional--Integral--Derivative (PID) feedback controller \cite{minorsky1922directional} tunes the system parameters while training. We focus on Chinese college students as the primary user group, providing a focused sample for testing the system.

To provide an initial consistency check, this study treats BlossomPsy as a black box and \textbf{compares} its results with the Chinese version of the \textbf{MBTI-M} assessment \cite{Cai2001MBTI}. The objective of BlossomPsy is to generate predictions that are consistent with MBTI-M while providing a more interactive assessment experience. This comparison should be interpreted as preliminary evidence of alignment with a commonly used reference scale rather than a complete psychometric validation.

\begin{figure}[tbp]
	\centering
	\includegraphics[width=0.5625\linewidth]{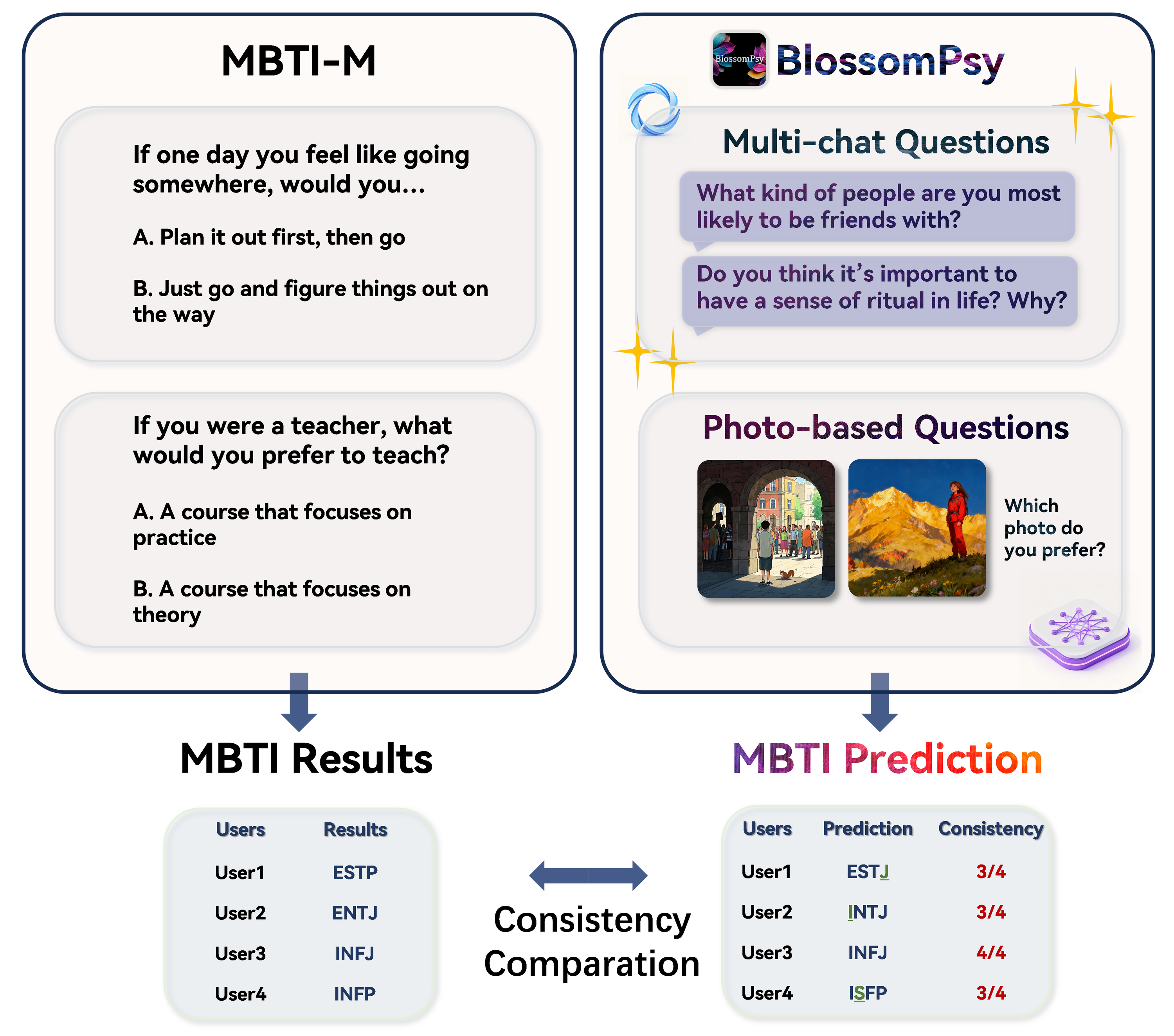}
    \caption{BlossomPsy, an MBTI Assessment System That Balances User Engagement with Predictive Consistency}
	\label{fig:motivation}
\end{figure}

In summary, our key contributions are as follows:
\begin{itemize}
    \item \textbf{AI-driven Hybrid Psychology Evaluation Framework:}
We design a novel multistage psychology evaluation pipeline that integrates multi-chat and photo-based questions to improve user engagement and predictive consistency; the questions are automatically modified according to users' responses.
    \item \textbf{Dynamic Confidence-driven Evaluation Mechanism:}
A MHC+mUCB evaluation system is proposed to dynamically estimate users' MBTI preferences. The framework uses PID feedback to tune confidence-scaling parameters during training.
\item \textbf{Iterative Supervised Visual Question Design:}
We introduce an iterative semi-automatic method to generate and screen photo-based MBTI items, aiming to preserve content alignment with traditional questions while enhancing engagement.
\item \textbf{Preliminary consistency assessment and user engagement:}
To evaluate assessment consistency, we compare results from BlossomPsy and MBTI-M, an established MBTI scale. User feedback was collected through a Likert scale to assess the interactive experience.
\end{itemize}
In Fig. \ref{fig:motivation}, an overview is provided to illustrate the structure of BlossomPsy and its consistency assessment.

\section{Related Work}

\subsection{AI-Based Personality Assessment}
Personality assessment has traditionally relied on standardized questionnaires such as the MBTI \cite{myers1962myers} and Big Five inventories \cite{gosling2003very}. Recent advances in NLP and LLMs have opened new avenues for automated personality prediction. \citet{peters2024large} studied whether LLMs can infer psychological dispositions from social media text, while \citet{maharjan2025psychometric} evaluated LLM embeddings for personality trait prediction. Recent work also highlights data and validity challenges in this area: \citet{li2024mbtibench} constructed MBTIBench to address noisy self-reported MBTI labels and population-distribution mismatch, and \citet{zhu2025personalityconversations} found that LLMs can show high test-retest reliability on real-world conversations while still exhibiting weak construct validity against Big Five self-reports. Interactive approaches have also emerged: \citet{li2025psydipersonalizedprogressivelyindepth} developed PsyDI, a multi-turn dialogue chatbot for MBTI assessment, and \citet{yang2024psychogat} combined psychological measurement with interactive fiction games. However, these systems typically lack systematic comparison with established scales or do not address the challenge of dynamically managing prediction confidence during interaction.

\subsection{MBTI: Utility and Psychometric Considerations}
The Myers-Briggs Type Indicator (MBTI) is one of the most widely used personality frameworks worldwide, with significant adoption in education, career counseling, and organizational settings \cite{myers1962myers}. However, its psychometric properties have been debated. \citet{pittenger1993measuring} raised concerns about test-retest reliability, and \citet{mccrae1989reinterpreting} showed that MBTI dimensions correlate with four of the Big Five personality traits (E/I with Extraversion, S/N with Openness, T/F with Agreeableness, J/P with Conscientiousness). These findings suggest that MBTI may reflect aspects of broader personality constructs, while its categorical interpretation should be treated cautiously. In this work, we adopt MBTI as the assessment framework for several practical reasons: its widespread familiarity among users enhances engagement, abundant labeled datasets exist for model training, and its four binary dimensions provide a clear structure for confidence-driven assessment. The proposed interaction and confidence-estimation framework is not specific to MBTI and could be adapted to other personality models such as the Big Five.

\subsection{Multi-Armed Bandits in Interactive Systems}
The multi-armed bandit (MAB) framework provides principled strategies for balancing exploration and exploitation under uncertainty \cite{auer2002finite}. MAB algorithms have been successfully applied in interactive systems including recommender systems, clinical trials, and adaptive testing. The Upper Confidence Bound (UCB) algorithm, which selects actions based on optimistic reward estimates, is particularly well-suited for scenarios where the system must decide which dimensions to probe further. In BlossomPsy, we adapt the UCB framework to personality assessment: each MBTI dimension is treated as an ``arm,'' and the algorithm identifies which dimension has the lowest prediction confidence, directing the conversation accordingly.

\subsection{LLM-Based Evaluation and Simulated Users}
The use of LLMs as evaluators and simulated participants has gained increasing acceptance in NLP and HCI research. \citet{zheng2023judging} established the LLM-as-a-Judge paradigm and reported high agreement with human evaluations in several settings. In personality research, \citet{serapio2023personality} showed that LLMs can exhibit stable and measurable personality patterns when given appropriate prompts, and \citet{bodrovza2024personality} further investigated the temporal stability of LLM personality responses. \citet{jiaqi2025comparative} compared LLM personality-like traits with human personality traits. Building on these findings, BlossomPsy employs LLMs as simulated test-takers to broaden MBTI profile coverage and stress-test the interaction pipeline, while human participants provide the primary real-user reference.

\section{Methodology}
\subsection{System Overview}
BlossomPsy is designed as a hybrid psychological assessment framework that combines multi-turn dialogue and adaptive photo-based items to balance engagement and predictive consistency. The system operates in three sequential stages:
\begin{enumerate}
    \item \textbf{Initial Profiling:} Users respond to a small set of background questions to initialize the model.
    \item \textbf{Multi-chatting:} LLM agents engage in personalized conversations with the user, guided by predictions from the evaluation module.
    \item \textbf{Prediction confirmation:} If prediction confidence is insufficient, the system dynamically inserts single-choice photo-based questions to refine the assessment.

\end{enumerate}

\begin{figure}[tbp]
    \centering
    \includegraphics[width=1\linewidth]{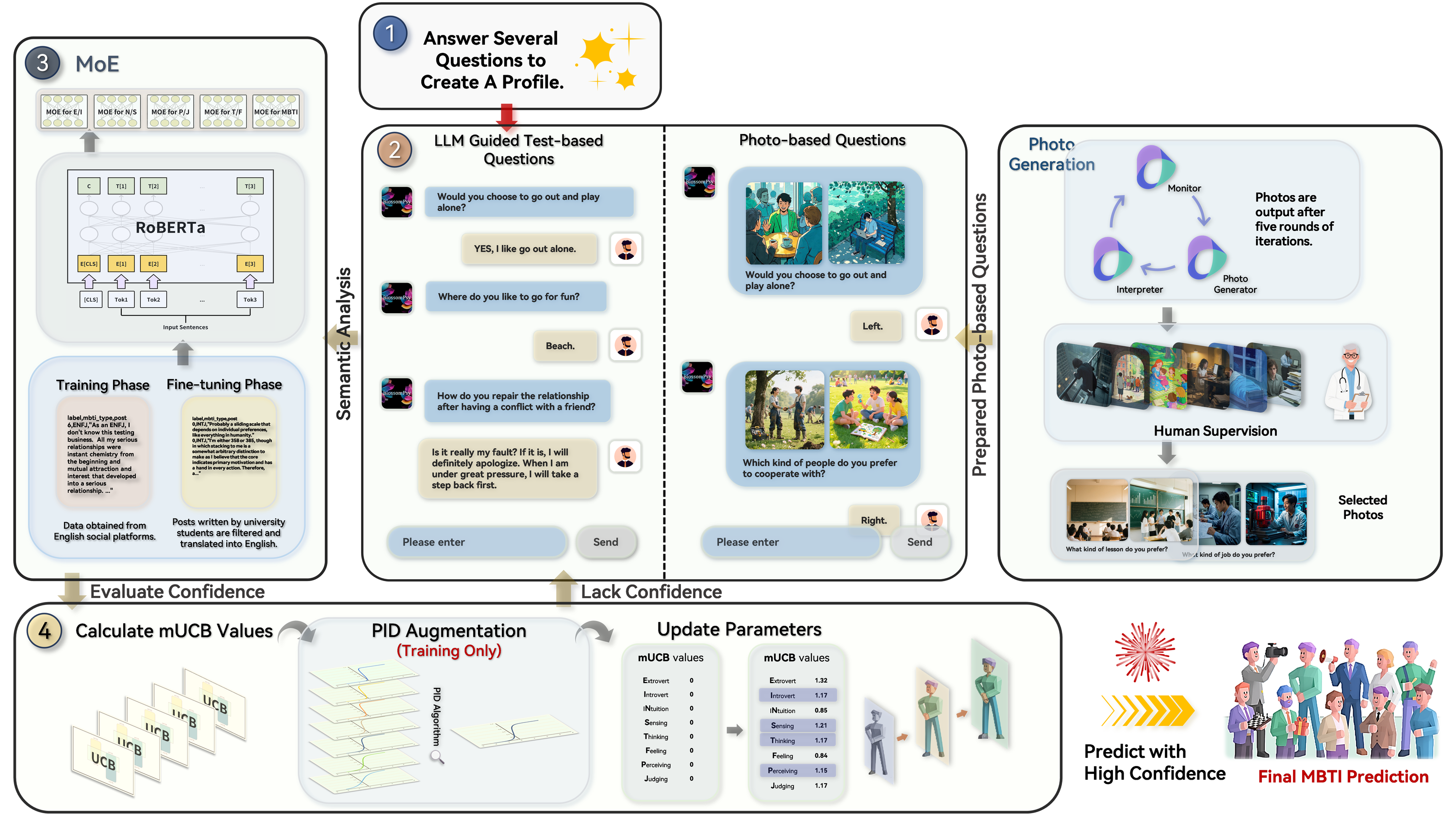}
    \caption{System Architecture of BlossomPsy, Including Dialogue, Photo-based Questions, and Prediction
 Modules}
    \label{fig:pipeline}
\end{figure}

The overall structure is shown in Fig. \ref{fig:pipeline}, including the dialogue and the photo-based prediction module, a prediction mechanism that integrates the confidence-based evaluation mechanism and the photo generation pipeline.
\FloatBarrier
\subsection{Agent for Multi-chatting}
\begin{figure}[tbp]
    \centering
    \includegraphics[width=0.75\linewidth]{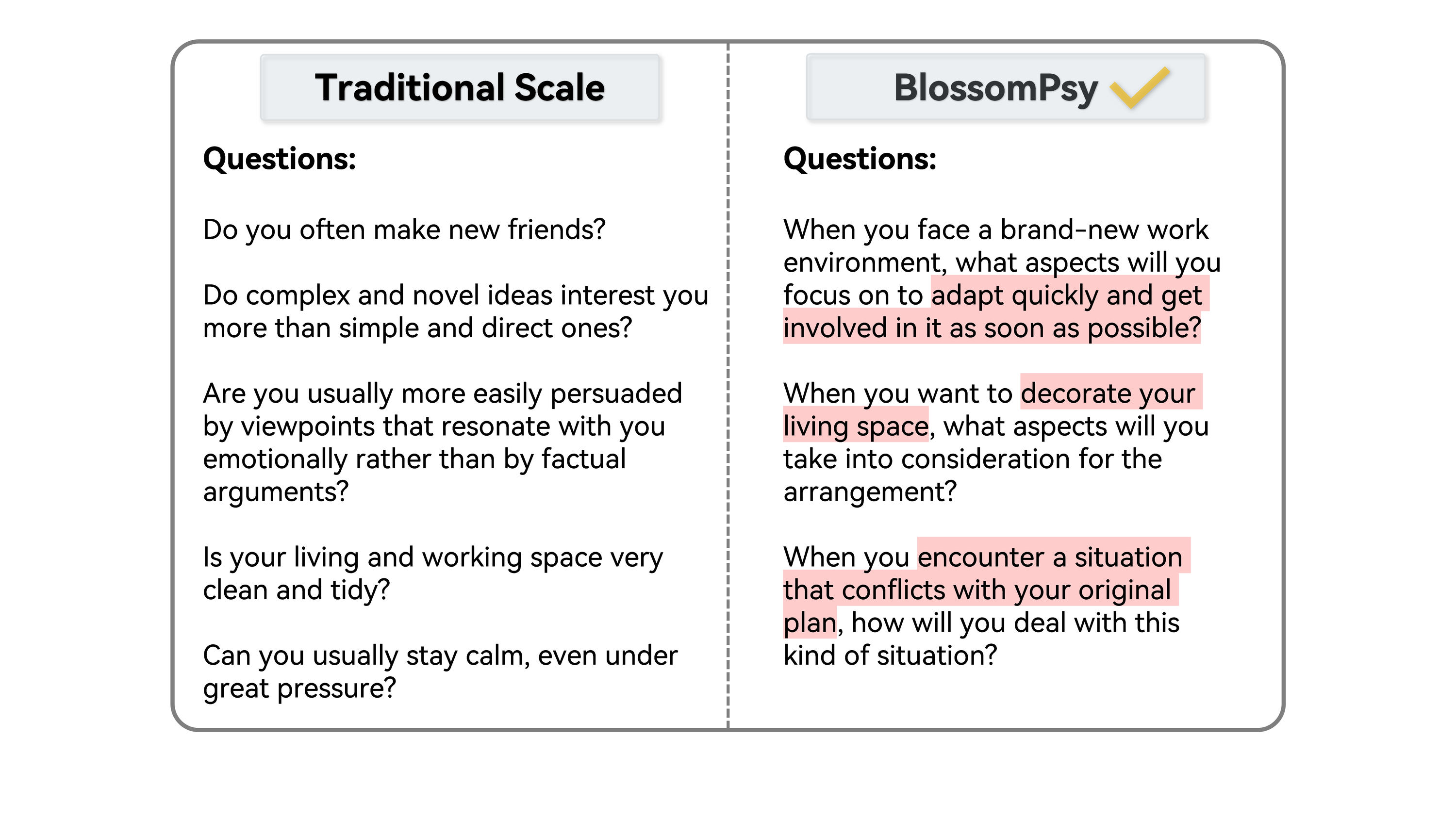}
    \caption{Example of Questions from MBTI-M (Chinese version) and BlossomPsy.}
    \label{fig:comparisom example}
\end{figure}

The multi-chat agent consists of two cooperating LLMs from ByteDance \cite{doubao2025} that drive multi-turn conversations with the user. In each round, the user responds in text to a question generated by the LLM. To prevent the multi-chat model from repeatedly focusing on the same topic, one LLM is designated to lead the topic based on chat history and guidance from the evaluation module. The questions generated by the LLM are intended to encourage the user to elaborate, as the assessment mechanism predicts MBTI based on semantic styles. We carefully designed the prompts to encourage the LLM to generate open questions. Fig. \ref{fig:comparisom example} compares questions from MBTI-M (Chinese version) and BlossomPsy. In comparison, questions from BlossomPsy invite more personal answers. Prompts for the LLMs are provided in Fig. \ref{fig:chatting} in the Appendix.
\FloatBarrier

\subsection{Confidence Estimation via Multi-Head Classifier and Modified mUCB Algorithm}
\begin{figure}[tbp]
    \centering
    \includegraphics[width=0.5\linewidth]{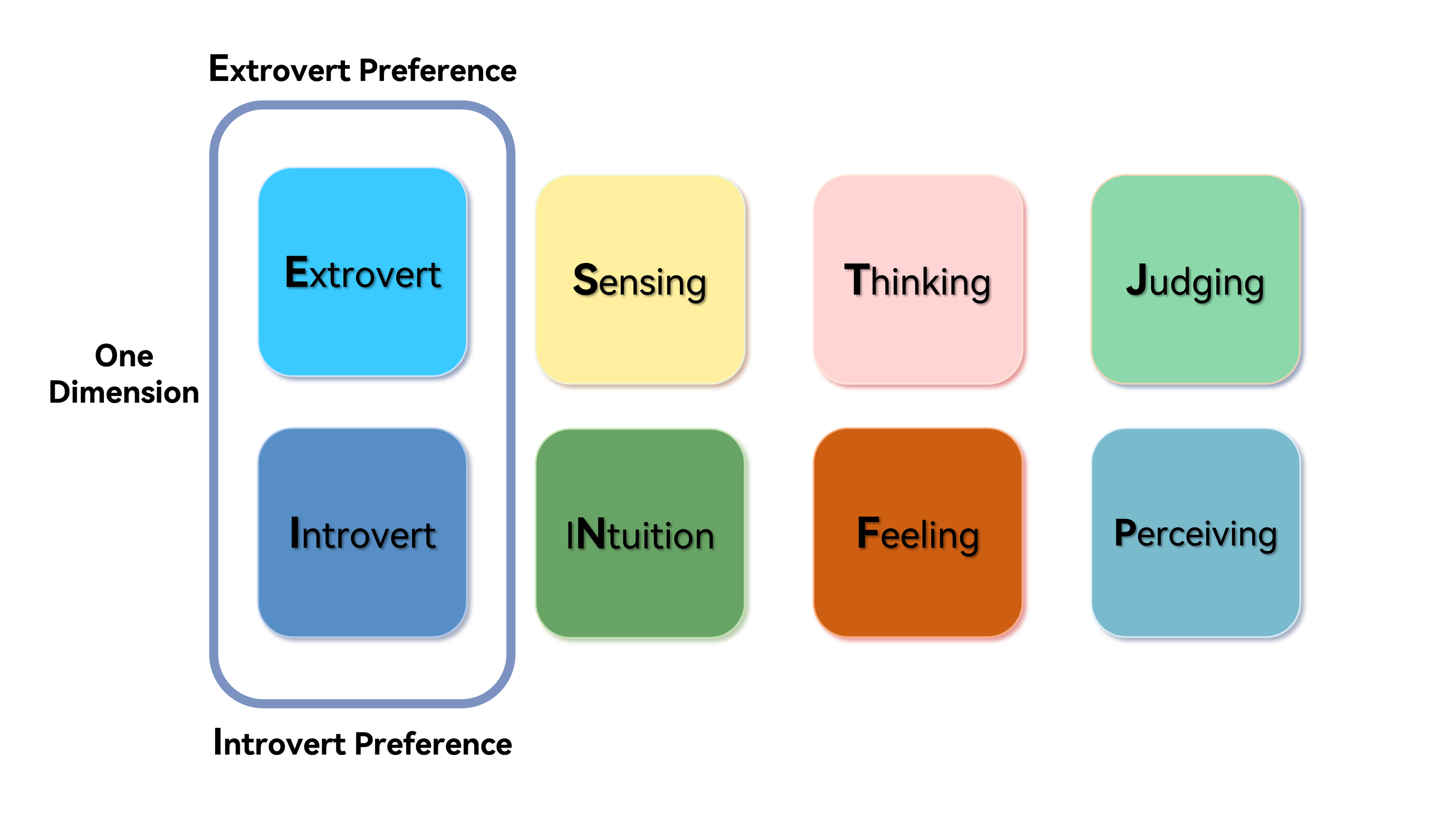}
    \caption{The Four MBTI Dimensions (E/I, S/N, T/F, J/P) Used in Classification. }
    \label{fig:MBTI Type}
\end{figure}

The first part of the evaluation module is a Multi-Head Classifier (MHC) trained on a dataset based on Personality Cafe \cite{kaggle_dataset2025} and CPME \cite{zhou2024chinese}. Unlike a standard Mixture-of-Experts architecture with a learned gating mechanism \cite{jacobs1991adaptive}, the MHC uses a shared encoder with multiple parallel classification heads, each specializing in a distinct prediction task. All heads are active simultaneously during inference, as each addresses a different MBTI dimension. It comprises:
\begin{itemize}
    \item A pretrained \textbf{RoBERTa-base encoder}\cite{cui2020revisiting}, which transforms input text into a 768-dimensional embedding.
    \item \textbf{Five parallel multilayer perceptrons (MLPs),} which contain four binary classifiers that are used to predict the four MBTI dimensions individually: (\textbf{E}xtrovert/\textbf{I}ntrovert, \textbf{S}ensing/i\textbf{N}tuition, \textbf{T}hinking/\textbf{F}eeling, \textbf{J}udging/\textbf{P}erceiving) and one 16-class classifier to capture potential correlations between dimensions.
\end{itemize}
For better understanding, the 4 dimensions are defined according to the structure illustrated in Fig. \ref{fig:MBTI Type}
Then, to estimate personalities with historical information and guide the interaction, we introduce a \textbf{modified Upper Confidence Bound (mUCB)} algorithm, which is an adaptation of the \textbf{Upper Confidence Bound (UCB)} from the multi-armed bandit (MAB) framework. It operates on top of the MHC's logit outputs and makes predictions; the multi-chat agent then generates new questions according to the personality preference with the lowest prediction confidence, thereby iteratively refining the MBTI profile through dialogue. Details of the mUCB module are provided below:

\textbf{Logit Enhancement:} In the MHC, each classification head outputs a logit score for its corresponding personality dimension (e.g., Introvert: 0.72, Extrovert: 0.21). However, when the logits of two opposite preferences within the same MBTI dimension are nearly identical, mUCB struggles to discriminate between them effectively. We adopt a sigmoid-like enhancement function:
\begin{equation}
f(\mathbf{p_i}) = \frac{1}{1 + \alpha e^{-\beta\mathbf{p_i}}}
\end{equation}
where \(\mathbf{p_i}\) is the logit of dimension \(i\), and \(\alpha, \beta\) are parameters learned by the PID feedback control during training.

\textbf{mUCB and mLCB Formulation:} The enhanced scores are used to compute the mUCB value for each dimension \(i\). To avoid division by zero at the beginning of the interaction, each dimension is initialized with a pseudo-count \(T_i(0)=1\) and an initial enhanced score \(\hat{h}_i(0)=s_i(0)\), where \(s_i(t)=f(\mathbf{p_i}(t))\) denotes the observed enhanced score at round \(t\).
\begin{equation}
\mathrm{mUCB}_i(t)
  = \hat{h}_i(t-1)
  + \frac{1}{2}\sqrt{\frac{2\log(n)}{T_i(t-1)}}
\end{equation}

\begin{equation}
\mathrm{mLCB}_i(t)
  = \hat{h}_i(t-1)
  - \frac{1}{2}\sqrt{\frac{2\log(n)}{T_i(t-1)}}
\end{equation}

Here, \(T_i(t-1)\) is the number of times that dimension \(i\) has been selected before round \(t\), and \(\hat h_i(t-1)\) is the running mean of observed enhanced scores or rewards for dimension \(i\). After receiving a new observation \(s_i(t)\), the estimate is updated by:
\begin{equation}
    T_i(t)=T_i(t-1)+1,\quad
    \hat{h}_i(t)=\frac{T_i(t-1) \times \hat{h}_i(t-1)+s_i(t)}{T_i(t)}
\end{equation}

\begin{figure}[tbp]
    \centering
    \includegraphics[width=0.5\linewidth]{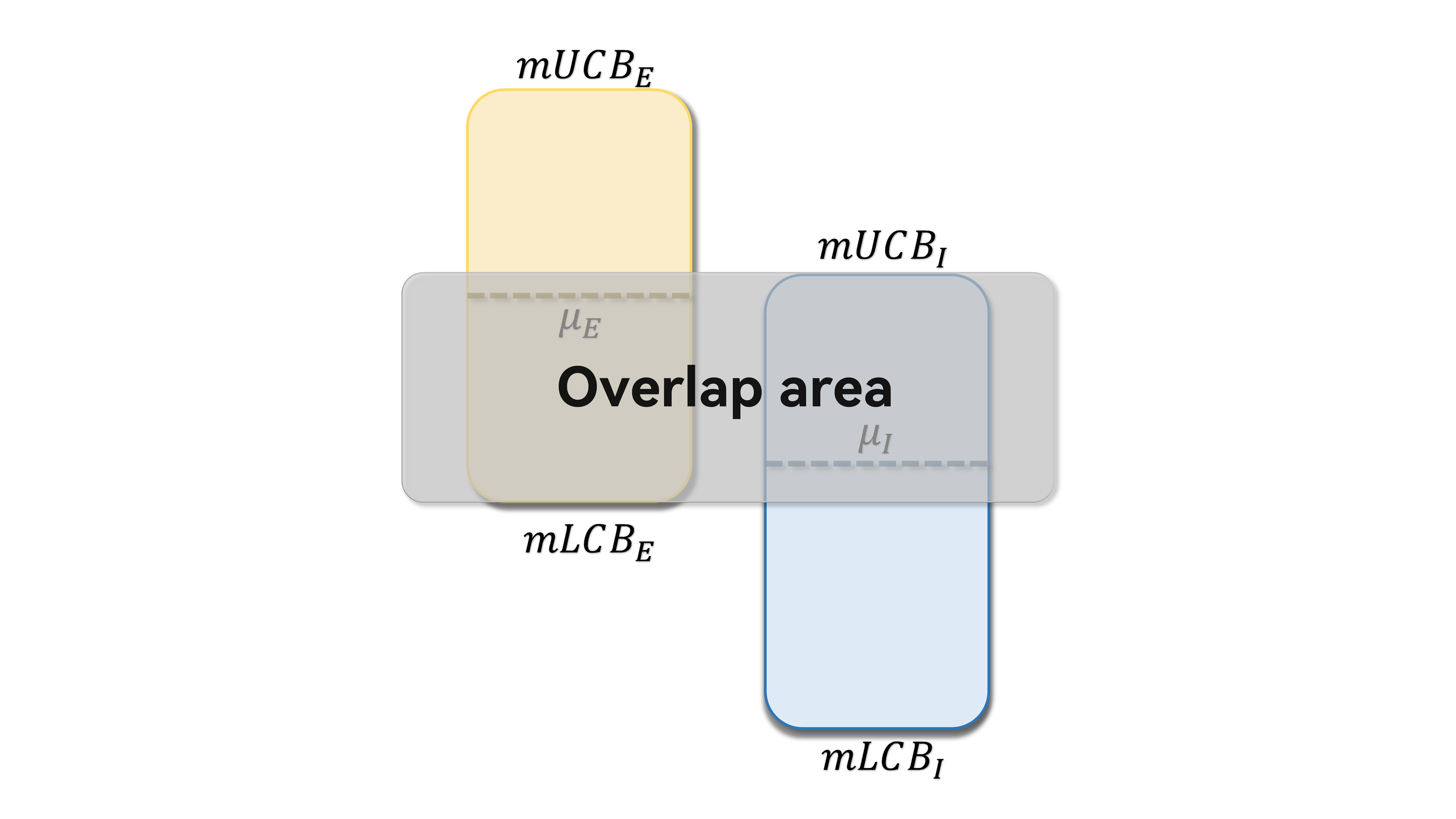}
    \caption{ Visualization of Overlap Rate Between mUCB and mLCB Intervals for a Given MBTI Class.}
    \label{fig:overlap}
\end{figure}

\textbf{Overlap Rate and Stopping Criterion:} Prediction confidence is quantified using the \textbf{overlap rate} between competing personality preferences in the same dimension. To handle cases where the confidence intervals do not overlap (resulting in a negative numerator), we apply a clamping operation:
\begin{equation}
    \mathbf{overlap~rate}=\max\!\left(0,\;\frac{\min(mUCB_i,\;mUCB_j)-\max(mLCB_{i},\;mLCB_{j})}{\max(mUCB_i,\;mUCB_j)-\min(mLCB_{i},\;mLCB_{j})}\right)
\end{equation}
When the overlap rate equals zero, the two preferences are well-separated, indicating high prediction confidence. Fig. \ref{fig:overlap} visualizes the overlap; the side length of the rectangle covered by the shaded region corresponds to the overlap area.

Then, a threshold of 0.6 is used; if the overlap exceeds this value after every five iterations, the system regards the prediction as a lack of certainty and triggers the module of photo-based questions.

\subsection{Visual Question Module for Low-Confidence Resolution }
\begin{figure}[tbp]
    \centering
    \includegraphics[width=0.75\linewidth]{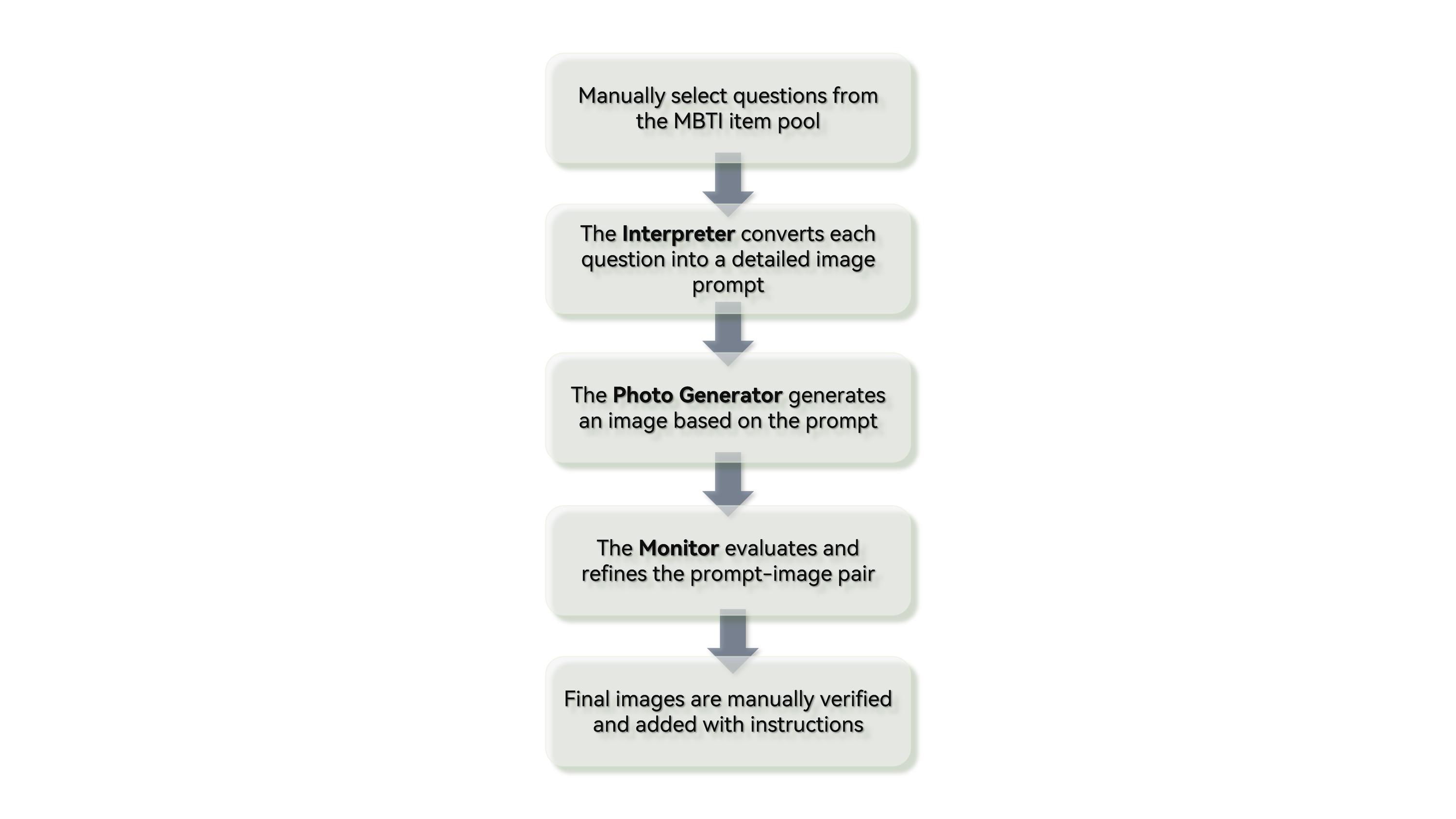}
    \caption{Workflow of the Visual Question Module for Low-confidence Fallback Evaluation. }
    \label{fig:photo pipeline}
\end{figure}

\begin{figure}[tbp]
    \centering
    \includegraphics[width=1\linewidth]{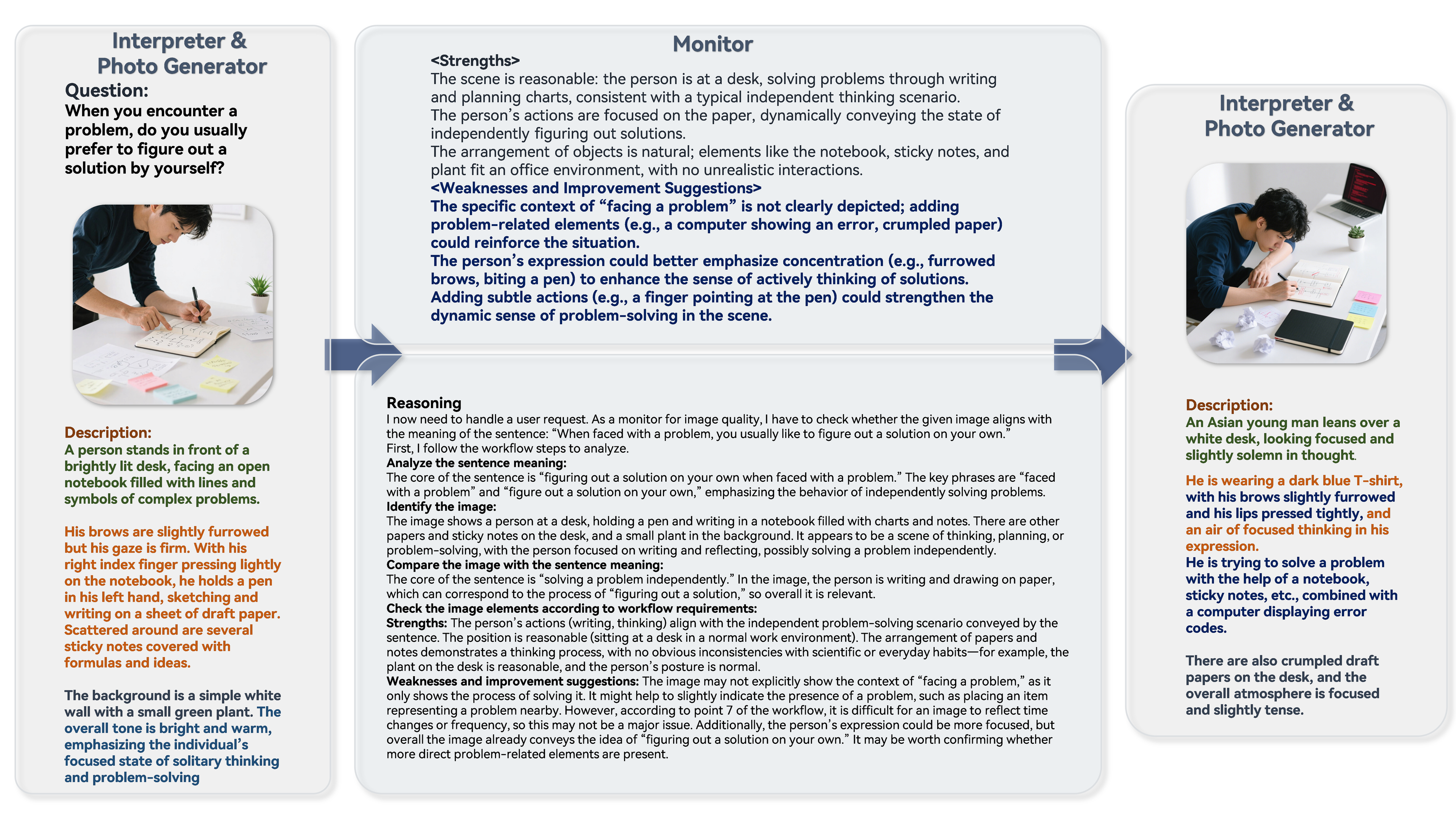}
    \caption{Cooperation of Different LLMs in the Photo Generation Module.}
    \label{fig:photo agents example}
\end{figure}

After activating the photo-based module, BlossomPsy presents the user with single-choice photo-based questions designed to probe the trait with low confidence. In addition to increasing prediction confidence, this module is intended to improve user engagement, as photos are generally more engaging and novel in assessment contexts.
As provided in Fig. \ref{fig:pipeline}, the photo generation procedure contains three LLMs and human supervision. The detailed pipeline is shown in Fig. \ref{fig:photo pipeline}. In consideration of efficiency and observed item quality, the pipeline is run for three iterations prior to human analysis. Sec. \ref{sec:photo-question-eval} provides more details of the related experiments.
To better illustrate the iteration process, Fig. \ref{fig:photo agents example} presents an example of the cooperation among the interpreter, the photo generator, and the monitor; After receiving a question, the interpreter generates a photo description consisting of an overview as well as descriptions of the host and the environment, which are shown in different colors. The monitor then analyzes the strength and weakness of the photo step by step under the guidance of the prompts. Finally, the photo generator modifies the photo description, the modifications are marked in orange. The prompts used for the three LLMs are provided in Fig. \ref{fig:photo agents} in the Appendix.
The following is some supplementary information for the photo-based questions;

\textbf{Validation Process:}
After the human supervision was conducted, a group of evaluators completed both the original textual question and its visual counterpart for validation. Only images with a response consistency of at least 66\% were retained.

\textbf{Personalization: }
We also consider individual differences. Previous studies indicate that people with different genders and personalities exhibit varying preferences for visual elements such as figures and colors \cite{photosupport,10.3233/AIS-160399}. For example, some females may dislike images dominated by aggressive male figures, and such preferences may bias their choices. To address this issue, we prepared several sets of photos, and the system changes figures in the photos based on the user's gender. This approach presents a feasible direction for future work------with enough statistical support, future research may further tailor photo backgrounds or color palettes to suit individual preferences.

\textbf{Integration with mUCB:}
When a user selects an option in a visual question, the system updates the reward count for the corresponding MBTI dimension:
\begin{equation}
T_i(t) = T_i(t-1) + 1
\label{eq:T construct}
\end{equation}

\begin{equation}
\hat{h}_i(t) = \frac{T_i(t-1) \times \hat{h}_i(t-1) + r_i(t)}{T_i(t)}
\label{eq:h construcct}
\end{equation}

In \eqref{eq:T construct} \eqref{eq:h construcct}, \(T_i(t)\) is the number of times that dimension \(i\) has been selected up to round \(t\), \(r_i(t)=1\) is the observed reward assigned to the selected dimension, and \(\hat{h}_i(t-1)\) is the running mean of previous observed rewards for dimension \(i\). This ensures that photo-based questions are incorporated into the mUCB pipeline, maintaining evaluation consistency.

\subsection{Training and Optimization}
To improve the robustness and consistency of the evaluation system in real-world usage, the training procedure is divided into three distinct stages: (1) training the MHC to predict MBTI types from user responses, (2) fine-tuning the MHC on target-domain samples, and (3) tuning the mUCB confidence transformation by PID feedback control.
\subsubsection{MHC Training Process}
In the first stage, the MHC was trained on the Personality Cafe MBTI dataset to learn to correctly predict the user's MBTI. This dataset comprised user-generated posts annotated with self-reported MBTI labels. Posts were filtered, segmented, and grouped by MBTI to ensure balanced representation among the 16 types. In this model, the pretrained RoBERTa-base model mapped each data sample to a 768-dimensional semantic representation. This representation was fed into five identical multilayer perceptrons (MLP). Four of these MLPs were binary classifiers, each responsible for predicting one MBTI dimension, while the fifth MLP performed a 16-class classification task, predicting the MBTI type as a whole. This MLP aimed to preserve potential inter-dimensional relationships.
Training was conducted using cross-entropy loss, with the F1 and AUC scores used to measure the model's performance. Dropout and weight decay were applied to prevent overfitting. Details of the model architecture and training parameters are illustrated in Tabs. \ref{tab:roberta-arch} and \ref{tab:train-config} in the Appendix.

In the second stage, the model underwent a fine-tuning process to further adapt to the target task. We filtered posts from CPME by identifying common phrases used by Chinese college students. The filtered posts were translated into English. During fine-tuning, the learning rate was set to \(2\times 10^{-6}\), while other parameters remained the same as in the training stage.
\subsubsection{Simulated Users behavior }
To tune the mUCB algorithm under noisy interaction conditions, synthetic users were generated using shuffled and partially corrupted samples. Each personality type's post set was mixed with 20\% posts randomly drawn from other MBTI types, introducing label noise and imitating real-world inconsistencies in user behavior. All the posts came from filtered CPME dataset.
\subsubsection{PID-Based Optimization for Confidence Estimation }
While the MHC produces logits for each dimension, the mUCB module requires confidence scores with a stable scale. In the third stage, a \textbf{Proportional--Integral--Derivative (PID)} controller was used to adjust the parameters \(\alpha\) and \(\beta\) in the sigmoid-like enhancement function \(f(\mathbf{p_i}) = \frac{1}{1 + \alpha e^{-\beta\mathbf{p_i}}} \). We used the simulated dataset during training. The PID controller updates \(\alpha\) and \(\beta\) according to Algorithm \ref{alg:update}.

% Triangle comment command aligned with the figure style
\newcommand{\TriComment}[1]{\Statex \hspace*{-1.5em} $\triangleright$~#1}
\begin{algorithm}[H]
\caption{Update Rule for $\alpha$ and $\beta$}
\label{alg:update}
\begin{algorithmic}[1]
\TriComment{$e_k$: error between $\mathrm{mUCB}(k)$ and the true reward $r(k)$}
\TriComment{$K_p = 1$ (proportional gain), $K_i = 0.02$ (integral gain), $K_d = 0.01$ (derivative gain).}
\TriComment{$r_{\text{correct}} = 0.5$ (reward for a correct decision), $r_{\text{incorrect}} = 0$ (reward for an incorrect decision).}
\TriComment{$\beta_k$: adaptive control variable at round $k$.}
\TriComment{$\alpha_k$: scaling parameter derived from $\beta_k$.}
\TriComment{$N = \lceil |\mathcal{D}_{\text{train}}| / 10 \rceil$; one PID update is performed every 10 training instances.}

\State Initialize $\alpha_0$, $\beta_0$
\For{$k = 1$ to $N$}
    \State $e_k \gets \mathrm{mUCB}(k) - r(k)$
    \State $\Delta \beta_k \gets K_p e_k + K_i \sum_{n=k-3}^k e_n + K_d(e_k - e_{k-1})$
    \State $\beta_{k+1} \gets \beta_k + \Delta \beta_k$
    \State $\alpha_{k+1} \gets \exp(\beta_{k+1} / 2)$
\EndFor
\end{algorithmic}
\end{algorithm}

\section{Experiments}
This section presents a series of experiments designed to evaluate the preliminary performance of BlossomPsy. Sec. \ref{sec:core-task-validation} reports consistency and user-feedback results involving both human participants and LLMs, while Sec. \ref{sec:component-analysis} analyzes the contribution of different system components and investigates parameter behavior.

\subsection{Experiment Setup}
A total of six experiments were conducted, each targeting a specific aspect of system evaluation:
\begin{itemize}
    \item \textbf{MBTI Consistency Evaluation:} Human volunteers and LLMs were invited to complete both BlossomPsy and the test from the Chinese version of the MBTI-M. Results were compared using accuracy, F1-score and Cohen's kappa.
    \item \textbf{User Experiments Evaluation: }Since user engagement is a core design goal, participants rated BlossomPsy on a five-point Likert scale across five dimensions: interactivity, immersion, clarity, interest, and satisfaction.
    \item \textbf{Evaluation of Photo-based Questions:} To screen photo-based questions, their responses were compared against their text-based counterparts. This was carried out by volunteers and LLMs. An agreement threshold of 0.66 was applied to distinguish candidate photo-based questions. To evaluate the photo generation workflow, a series of supplement experiments were conducted.
    \item \textbf{MHC Performance:} The MHC was tested on several datasets to assess its predictive performance. F1-score and accuracy are reported.
    \item \textbf{Ablation Study:} The contribution of PID, mUCB, and photo-based modules was evaluated by selectively removing components.
    \item \textbf{PID in parameter-finding process:} The convergence of parameters \(\alpha\) and \(\beta\) under PID control was analyzed to study whether PID feedback can find stable parameter settings.
\end{itemize}

\subsection{Core Task Validation}
\label{sec:core-task-validation}
\subsubsection{Evaluation Methodology: Human and LLM Participants}
\label{sec:evaluation-methodology}
The consistency evaluation involved 21 participants: 12 human volunteers (Chinese college students) and 9 LLM-simulated participants (Qwen \cite{bai2025qwen2} and Doubao series). The two groups serve complementary roles in the evaluation:

\begin{itemize}
    \item \textbf{Human participants} provide real-user reference data, as they can complete both BlossomPsy and the MBTI-M scale independently and reflect real participant behavior.
    \item \textbf{LLM participants} help with scalability testing and broader coverage of MBTI profiles. Recent research suggests that LLMs can exhibit stable, measurable personality patterns under role-playing prompts \cite{serapio2023personality,bodrovza2024personality,jiaqi2025comparative}, and the LLM-as-a-Judge paradigm has reported high agreement with human evaluations in related tasks \cite{zheng2023judging}.
\end{itemize}

\subsubsection{MBTI consistency evaluation}
All 21 participants completed both BlossomPsy and the Chinese version of the MBTI-M scale. The results were compared on a dimension-by-dimension basis. Tab. \ref{tab:total result} shows the percentage of test-takers whose results matched BlossomPsy completely or partially, and Tab. \ref{tab:kappa} reports accuracy, F1-scores and Cohen's kappa across each MBTI dimension.

\begin{table}[ht]
    \centering
    \caption{BlossomPsy's Results Compared with the MBTI-M (Chinese Version) Test (Used as the Reference)}
    \label{tab:total result}
    \begin{tabular}{lc}
        \toprule
        Matched Dimensions (out of 4) & Percentage of Test-takers \\
        \midrule
        Match 4 dimensions & 33.33\% \\
        Match 3 dimensions & 42.86\% \\
        Match 2 dimensions & 14.29\% \\
        Match 1 dimension  &  9.52\% \\
        Match 0 dimensions &  0.00\% \\
        \bottomrule
    \end{tabular}
\end{table}

\begin{table}[ht]
    \centering
    \caption{BlossomPsy's Results Across Dimensions. For F1-score, the first-listed personality preference in each dimension is treated as the positive class.}
    \label{tab:kappa}
    % Keep all three resizebox arguments closed, with no extra characters after percent.
    \resizebox{\textwidth}{!}{%
        \begin{tabular}{cccccc}
            \toprule
            Metric & Extrovert/Introvert & Sensing/iNtuition & Thinking/Feeling & Judging/Perceiving & Average \\
            \midrule
            Accuracy & 0.76 & 0.67 & 0.95 & 0.76 & 0.79 \\
            F1-score & 0.79 & 0.72 & 0.92 & 0.78 & 0.81 \\
            Cohen's Kappa & 0.63 & 0.5 & 0.9 & 0.62 & 0.66 \\
            \bottomrule
        \end{tabular}
    }
\end{table}
The results show that over 75\% of participants had at least three dimensions matched with MBTI-M. The Cohen's kappa values range from 0.50 to 0.90 across dimensions, indicating moderate to almost-perfect agreement \cite{wongpakaran2013comparison}. Notably, the Thinking/Feeling dimension achieved the highest agreement ($\kappa = 0.90$), while the Sensing/iNtuition dimension showed only moderate agreement ($\kappa = 0.50$). This variation is consistent with the literature, as the S/N dimension is known to be harder to assess through text-based methods due to its abstract, cognitive nature. These results provide a preliminary reference point for future studies on interactive MBTI systems.

In Tab. \ref{tab:kappa}, the accuracy of a specific dimension is calculated as follows: let T be the number of correct predictions (from both preferences), and F the number of incorrect predictions. Then,
\begin{equation}
    \text{Accuracy of a specific dimension} = \frac{T}{T+F}
\end{equation}

During testing, BlossomPsy generated predictions based on the value of mUCB. Readers may refer to Fig. \ref{fig:mUCB_trends} in the Appendix for an example of the trends of mUCB in real-world settings.
\subsubsection{User Experiments Evaluation}
\begin{figure}[tbp]
    \centering
    \includegraphics[width=0.5\linewidth]{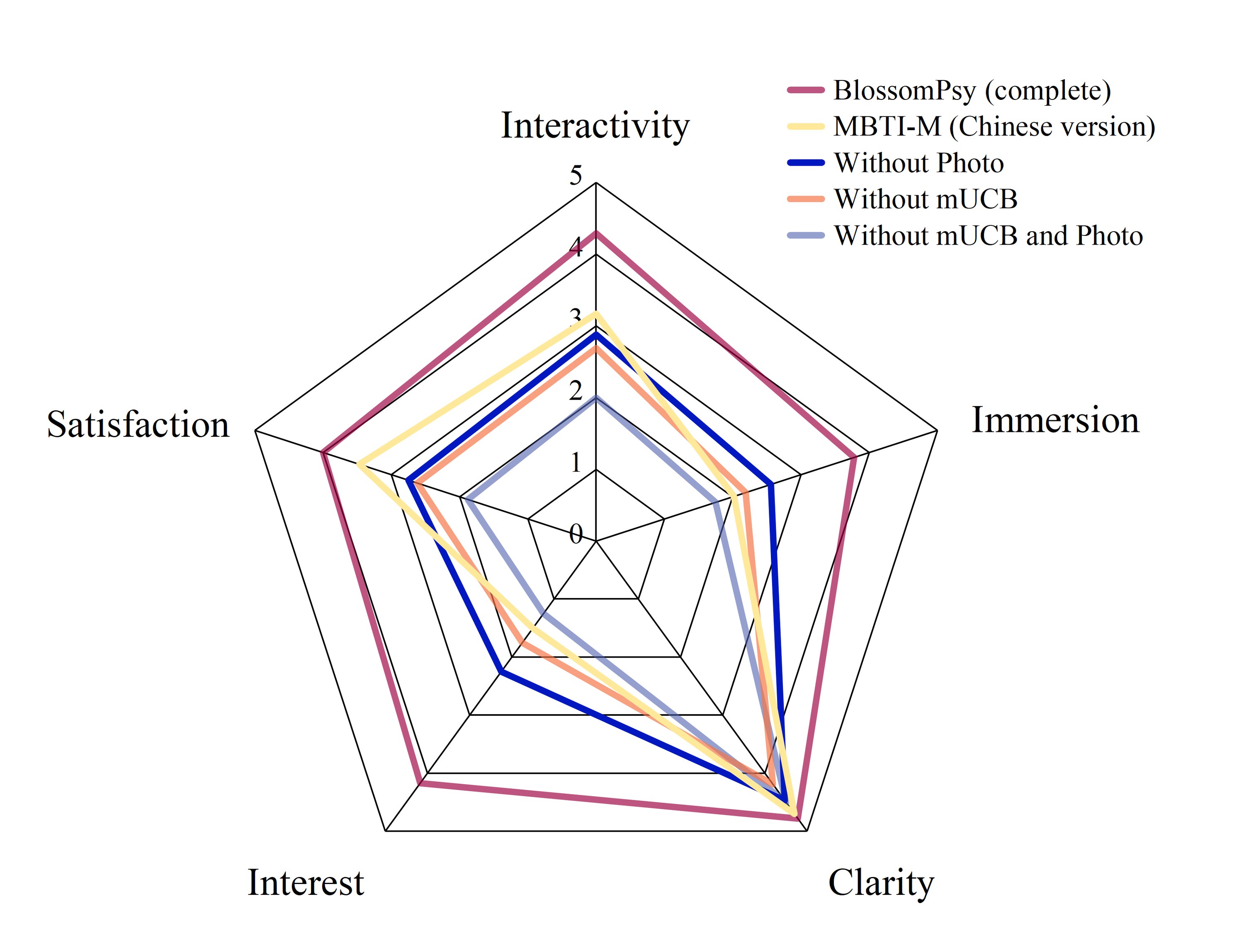}
    \caption{Test-takers' Ratings Across Five Evaluation Metrics (Interactivity, Immersion, Clarity, Interest, and Satisfaction).}
    \label{fig:vote}
\end{figure}

After completing the BlossomPsy test, participants rated the system on five dimensions (Fig.  \ref{fig:vote}). The results show that clarity was consistently maintained across different conditions. Compared with the Chinese version of the MBTI-M scale, BlossomPsy received higher user ratings in interactivity and interest. Test-takers also reported higher levels of satisfaction and immersion, suggesting that the combination of the multi-chat module and photo-based questions can improve the assessment experience.

\subsubsection{Evaluation of Photo-based Questions}
\label{sec:photo-question-eval}
To generate high-quality photos, we first selected text-based questions from the MBTI-M (Chinese version) scale according to the following rules, taking into account what information can be easily conveyed through photos:
\begin{itemize}
    \item The selected questions should ensure comprehensive coverage of the personality dimensions.
    \item Questions with clear, easily depicted scenes are given high priority.
    \item  A text-based question should be suitable to be transformed into a pair of photos, allowing test-takers with different personalities to choose the photo they prefer.  For example, the item \textit{``You rarely worry about whether you can make a good impression on people you meet.''} is not ideal for visual representation as it focuses on situations that span a long period of time.

\end{itemize}

\begin{figure}[tbp]
    \centering
    \includegraphics[width=0.75\linewidth]{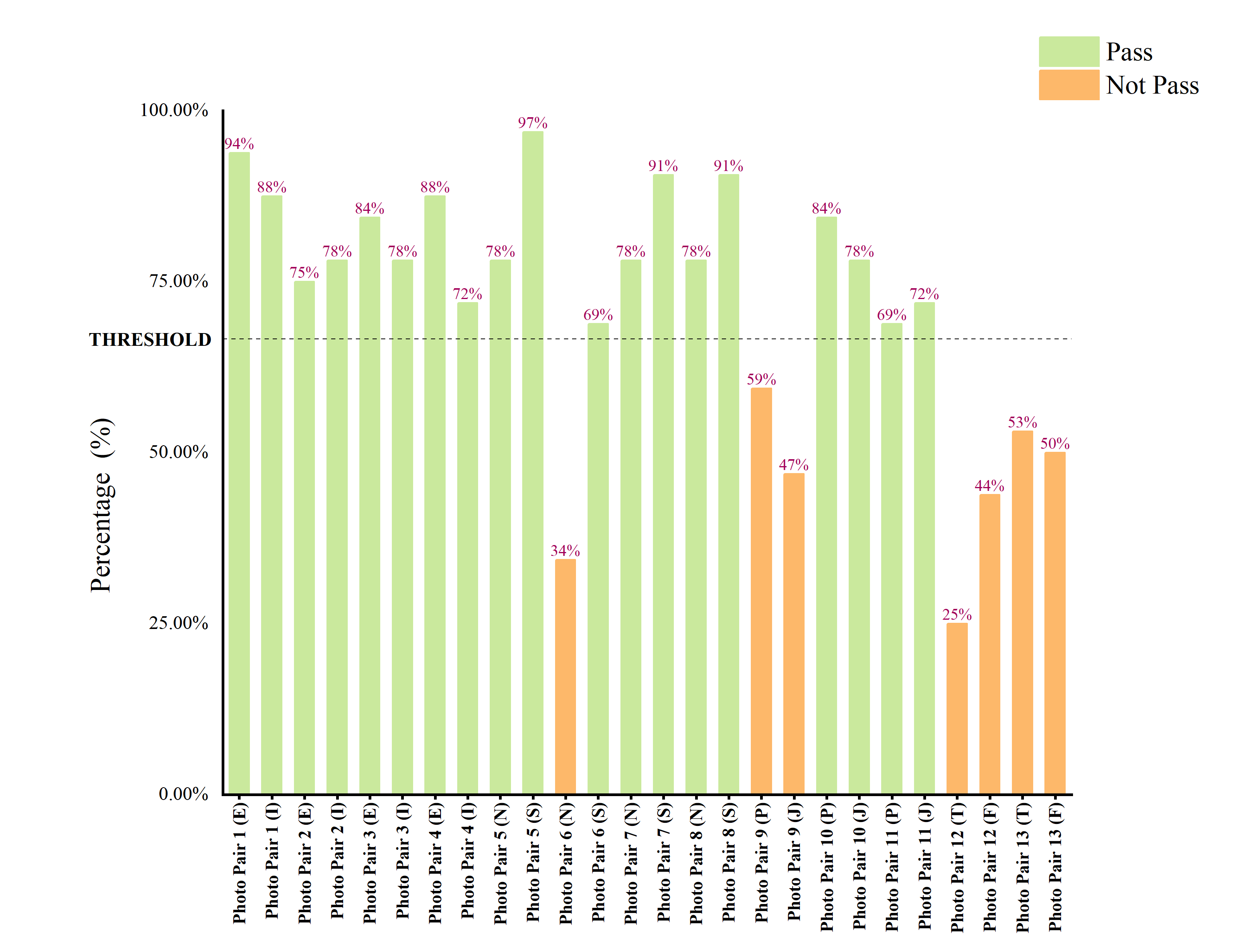}
    \caption{Results of Candidate Photo-based Questions, Showing the Percentage of Agreement with Text-based Counterparts.}
    \label{fig:photo vote 1}
\end{figure}

Following these rules, we selected 13 text-based questions and supervised the LLM to generate 13 corresponding pairs of photos. For validation, 10 human volunteers completed both the text-based and the photo-based questions. To increase coverage across MBTI profiles, the Qwen LLM was additionally used to generate 40 simulated evaluations. In this setting, each photo pair received a total of 50 evaluations. If the convergence fell below expectations, the photos were adapted and re-tested. As shown in Fig. \ref{fig:photo vote 1}, the threshold was set at 66\%; a photo was selected only if its \textbf{paired photo} exceeded this threshold. Based on this criterion, 9 photo pairs were finally retained. Some pairs of photos did not reach the threshold after several adaptations, so these photos were abandoned.
To examine the effect of the monitor, human supervision and the number of iterations, we employed the same Qwen LLM to simulate 16 test-takers with different MBTI types to complete the evaluation experiment. The result is shown in Tab. \ref{tab:HS} (\textit{HS} is the abbreviation for \textit{Human Supervision}), where each module generated 13 pairs of photos, and the acceptance rate as well as the average convergence rate are reported.
\begin{table}[htbp]
    \centering
    \caption{Comparison Between Different Versions of the Photo Generation Module}
    \label{tab:HS}
    \small
    \setlength{\tabcolsep}{3pt}
    \begin{tabularx}{\textwidth}{
        >{\raggedright\arraybackslash}X
        >{\centering\arraybackslash}p{3.2cm}
        >{\centering\arraybackslash}p{3.2cm}
    }
    \toprule
    Module & Acceptance Rate (Threshold=0.66)& Average Convergent Rate \\
    \midrule
    Original (3 Iterations)& 0.692& 0.746 \\
    w/o HS (5 Iterations)& 0.308& 0.534 \\
    w/o HS (4 Iterations)& 0.270& 0.451 \\
    w/o HS (3 Iterations)& 0.308& 0.581 \\
    w/o HS (2 Iterations)& 0.231& 0.475 \\
    w/o HS (1 Iteration)& 0.231& 0.442 \\
    w/o HS, w/o Monitor (3 Iterations)& 0.308& 0.505 \\
    \bottomrule
    \end{tabularx}
\end{table}

According to Tab. \ref{tab:HS}, human supervision was associated with a higher acceptance rate in this setting. The monitor was helpful in stabilizing the convergence rate. Increasing the number of iterations also improved the acceptance rate up to three iterations. Based on this observation, the photo generation module incorporated both the monitor and human supervision, with the number of iterations set to three. For more details on the modules presented in Tab. \ref{tab:HS}, readers may refer to Fig. \ref{fig:all} in the Appendix.
\subsection{Component Analysis and Parameter Behavior  }
\label{sec:component-analysis}
\subsubsection{MHC Performance}
The MHC, which integrates a pretrained RoBERTa encoder with five parallel MLP classification heads, was evaluated on multiple datasets. To the best of our knowledge, there is currently a lack of MBTI datasets specifically sampled from college students. To address this gap, we filtered target posts from the CPME dataset using words commonly used by college students. These posts were translated, divided, and independently employed during the fine-tuning and testing stages. In addition, we compared the MHC against other widely used algorithms in classification tasks. As shown in Tab. \ref{tab:XGBoost}, the MHC with RoBERTa achieved strong performance when trained on the Personality Cafe dataset and fine-tuned on the filtered CPME subset.
\begin{table}[htbp]
    \centering
    \caption{Comparison of Model Performance Across Different Datasets and Personality Dimensions.}
    \label{tab:XGBoost}
    \resizebox{\textwidth}{!}{
    \begin{tabular}{lcccccccccccccccc}
        \toprule
        \multirow{2}{*}{Dataset}& \multicolumn{4}{c}{MHC (RoBERTa)}
        & \multicolumn{4}{c}{MHC (BERT)}
        & \multicolumn{4}{c}{XGBoost}
        & \multicolumn{4}{c}{SVM} \\
        \cmidrule(lr){2-5} \cmidrule(lr){6-9} \cmidrule(lr){10-13} \cmidrule(lr){14-17}
        & I/E & N/S & T/F & P/J
        & I/E & N/S & T/F & P/J
        & I/E & N/S & T/F & P/J
        & I/E & N/S & T/F & P/J \\
        \midrule
        Filtered CPME& 0.663 & 0.645 & 0.692 & 0.633
             & 0.679 & 0.667 & 0.703 & 0.656
             & 0.535 & 0.564 & 0.621 & 0.580
             & 0.552 & 0.532 & 0.581 & 0.540 \\
        CPME + Fine-tuning
             & 0.659 & 0.646 & 0.689 & 0.631
             & 0.659 & 0.641 & 0.688 & 0.627
             & \multicolumn{4}{c}{N/A}
             & \multicolumn{4}{c}{N/A} \\
        Personality Cafe
             & 0.488 & 0.502 & 0.514 & 0.513
             & 0.496 & 0.502 & 0.503 & 0.510
             & 0.537 & 0.487 & 0.536 & 0.490
             & 0.650 & 0.700 & 0.615 & 0.576 \\
        Personality Cafe + Fine-tuning
             & \textbf{0.838}& \textbf{0.920} & \textbf{0.797} & \textbf{0.759}
             & 0.755 & 0.859 & 0.672 & 0.635
             & \multicolumn{4}{c}{N/A}
             & \multicolumn{4}{c}{N/A} \\
        \bottomrule
    \end{tabular}
    }
\end{table}

In addition, we trained the MHC on Personality Cafe and compared it with existing approaches. The results (Tab. \ref{tab:friends}) show that BlossomPsy's MHC achieves competitive performance, with balanced accuracy across all MBTI dimensions. Tab. \ref{tab:XGBoost} reports our controlled model-selection results across training and fine-tuning settings, whereas Tab. \ref{tab:friends} follows the percentage-style reporting used in prior Personality Cafe studies for contextual comparison. The BlossomPsy row in Tab. \ref{tab:friends} is therefore marked as a contextual comparison setting and should not be read as the same experimental setting as the Personality Cafe + Fine-tuning row in Tab. \ref{tab:XGBoost}. The table should not be read as a direct leaderboard because the compared studies differ in datasets, preprocessing, augmentation strategies, metrics, and evaluation settings.

\begin{table}[htbp]
\centering
\caption{Contextual Comparison of BlossomPsy's MHC with Existing Models}
\label{tab:friends}
\scriptsize
\setlength{\tabcolsep}{2pt}
\renewcommand{\arraystretch}{1.25}

\begin{tabular}{
    >{\raggedright\arraybackslash}p{2.8cm}
    >{\raggedright\arraybackslash}p{1.9cm}
    >{\raggedright\arraybackslash}p{2.3cm}
    >{\raggedright\arraybackslash}p{2.3cm}
    *{4}{>{\centering\arraybackslash}p{0.72cm}}
}
\toprule
\textbf{Paper / Work} & \textbf{Dataset} & \textbf{Augmentation} & \textbf{Method} & \textbf{I/E} & \textbf{S/N} & \textbf{F/T} & \textbf{J/P} \\
\midrule
M. H. Amirhosseini et al. & Personality Cafe & No & XGBOOST, RNN & 78\% & 86\% & 72\% & 66\% \\
V. G. dos Santos et al.   & Personality Cafe & No & BERT, LSTM   & 94\% & 91\% & 89\% & 91\% \\
P. Canbay                 & Personality Cafe & No & BERT         & 66\% & 70\% & 72\% & 66\% \\
\textbf{MHC in BlossomPsy (contextual setting)} & Personality Cafe & Over/under-sampling; percentage-style report & RoBERTa, MHC & 84\% & 87\% & 86\% & 81\% \\
X. Wu et al.              & Reddit & No & BERT & 69\% & 77\% & 68\% & 61\% \\
A. R. Julianda et al.     & Reddit & Over/under-sampling & NB, SVM, LR, BERT & \multicolumn{4}{c}{Total accuracy: 34\%} \\
\bottomrule
\end{tabular}

\end{table}

\subsubsection{Ablation Study}
\begin{figure}[tbp]
    \centering
    \includegraphics[width=0.75\linewidth]{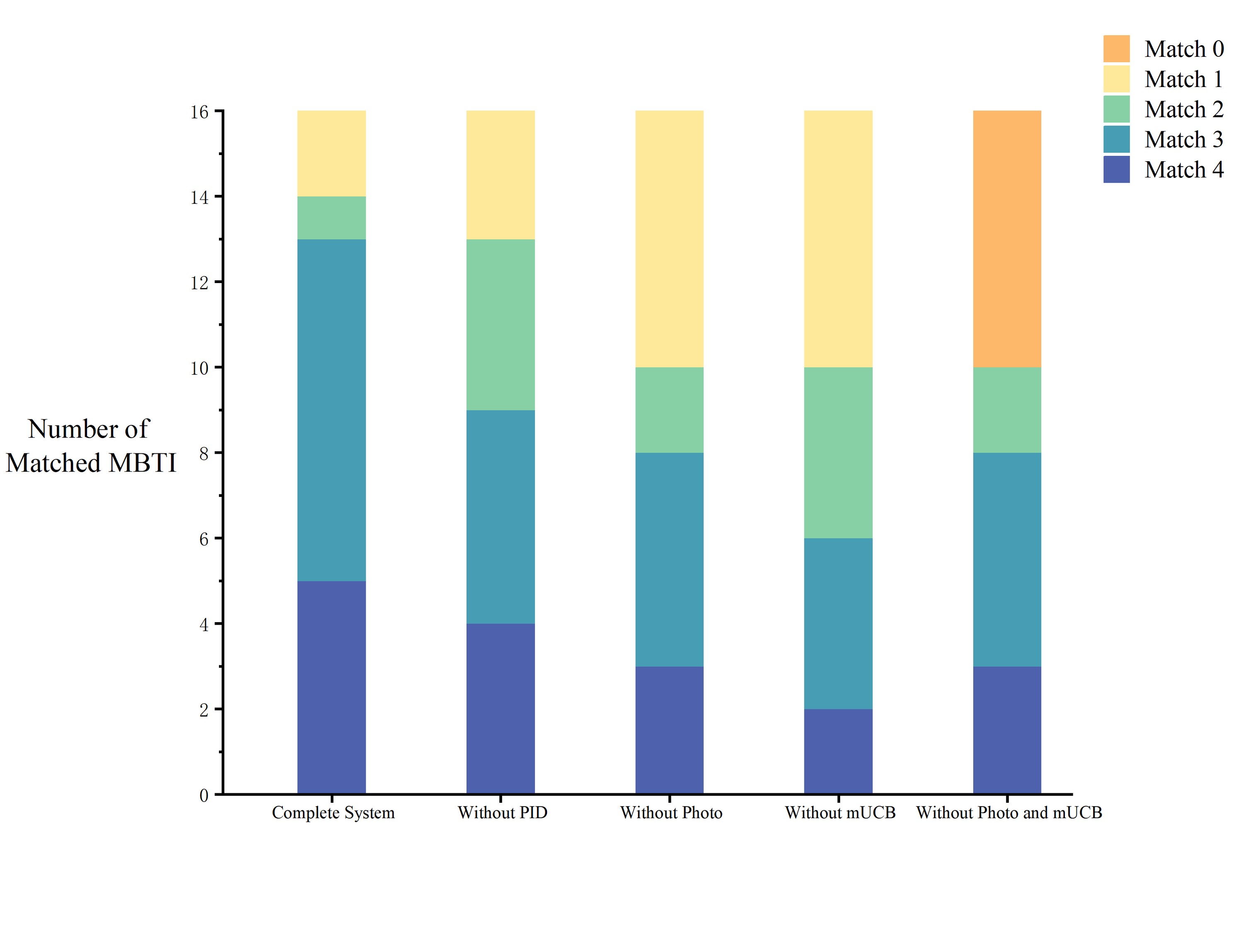}
    \caption{Performance Comparison of Different System Versions in the Ablation Study.}
    \label{fig:match}
\end{figure}

In this experiment, we focused on evaluating the effectiveness of the PID, mUCB algorithm and the photo-based questions. For each version, the LLM independently simulated volunteers representing 16 different MBTI types, and the results are shown in Fig. \ref{fig:match}.
Overall, the complete version performed similarly to the standard MBTI consistency evaluation. In this ablation setting, PID and photo-based questions were associated with higher prediction accuracy, particularly by increasing the proportion of highly matched responses. The mUCB algorithm helped stabilize the results by reducing completely incorrect predictions. In summary, the ablation results suggest that BlossomPsy can integrate multi-modal information in a way that improves prediction consistency and engagement in this experimental setting.

\subsubsection{PID in parameter-finding process}
We focus on how to determine suitable values of \(\alpha\) and \(\beta\) using the PID algorithm. These two parameters play a crucial role in the Sigmoid-like function, which is applied to augment the logit output of the MHC. We designed the experiment by placing the MHC+mUCB module in a simulated environment, where the PID algorithm guided the adjustment of these parameters.

\begin{figure}[tbp]
    \centering
    \begin{subfigure}[t]{0.48\linewidth}
        \centering
        \includegraphics[width=\linewidth]{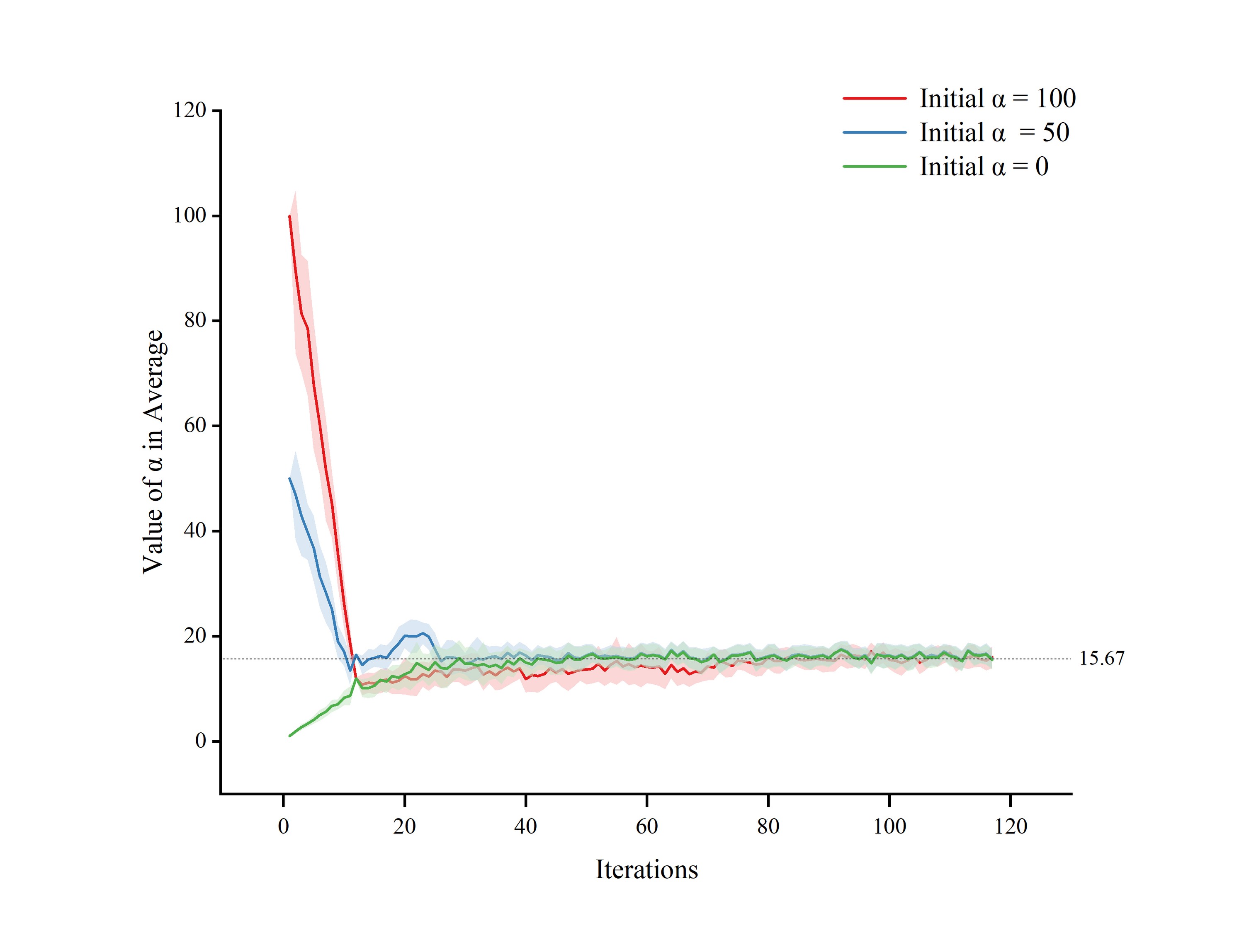}
        \caption{Trend of \(\alpha\) Parameter Adjustment under PID Control}
        \label{fig:alpha}
    \end{subfigure}
    \hfill
    \begin{subfigure}[t]{0.48\linewidth}
        \centering
        \includegraphics[width=\linewidth]{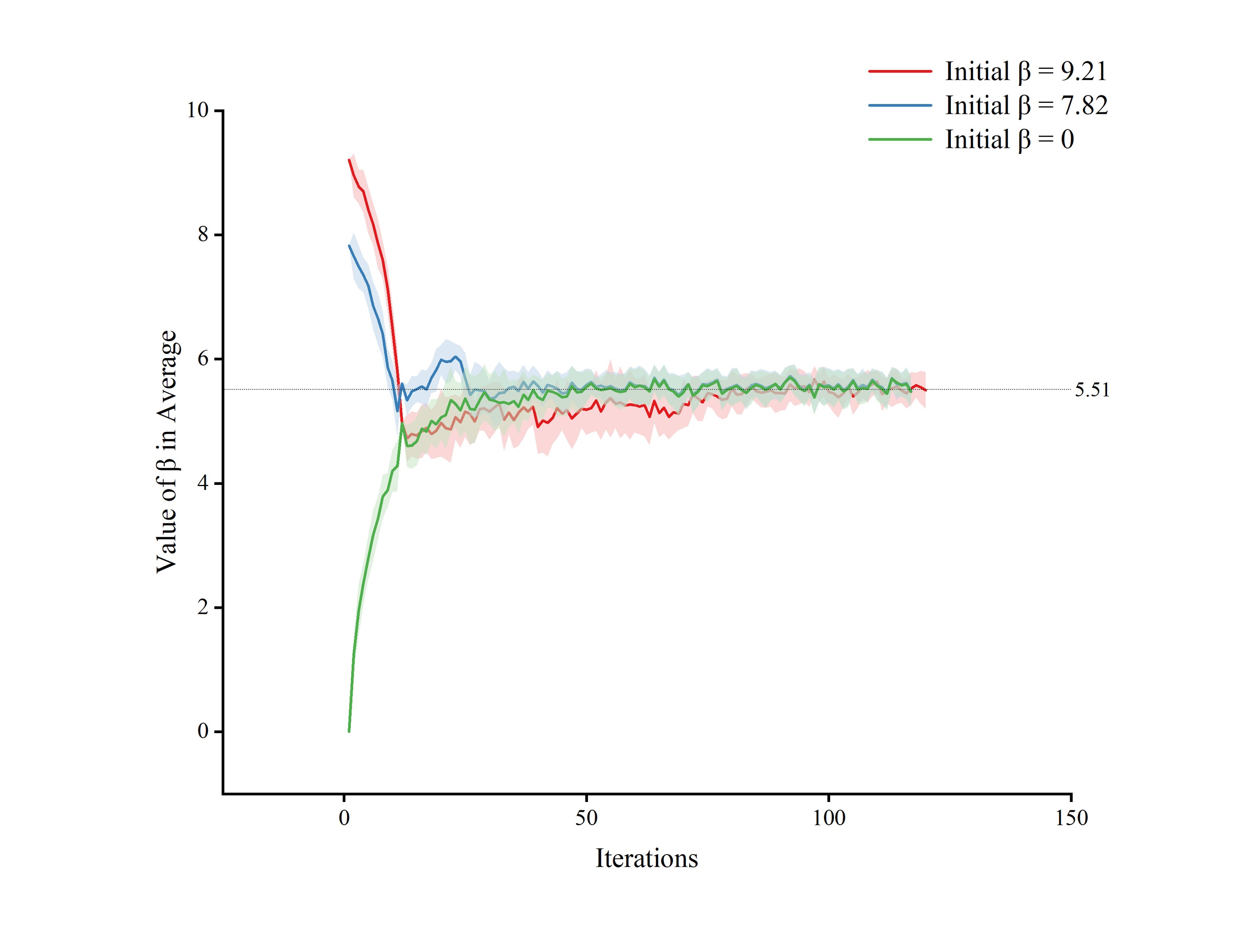}
        \caption{Trend of \(\beta\) Parameter Adjustment under PID Control  }
        \label{fig:beta}
    \end{subfigure}
    \caption{Trend of Parameters Adjustment under PID Control}
    \label{fig:alpha_beta}
\end{figure}

\begin{figure}[tbp]
    \centering
    \includegraphics[width=0.75\linewidth]{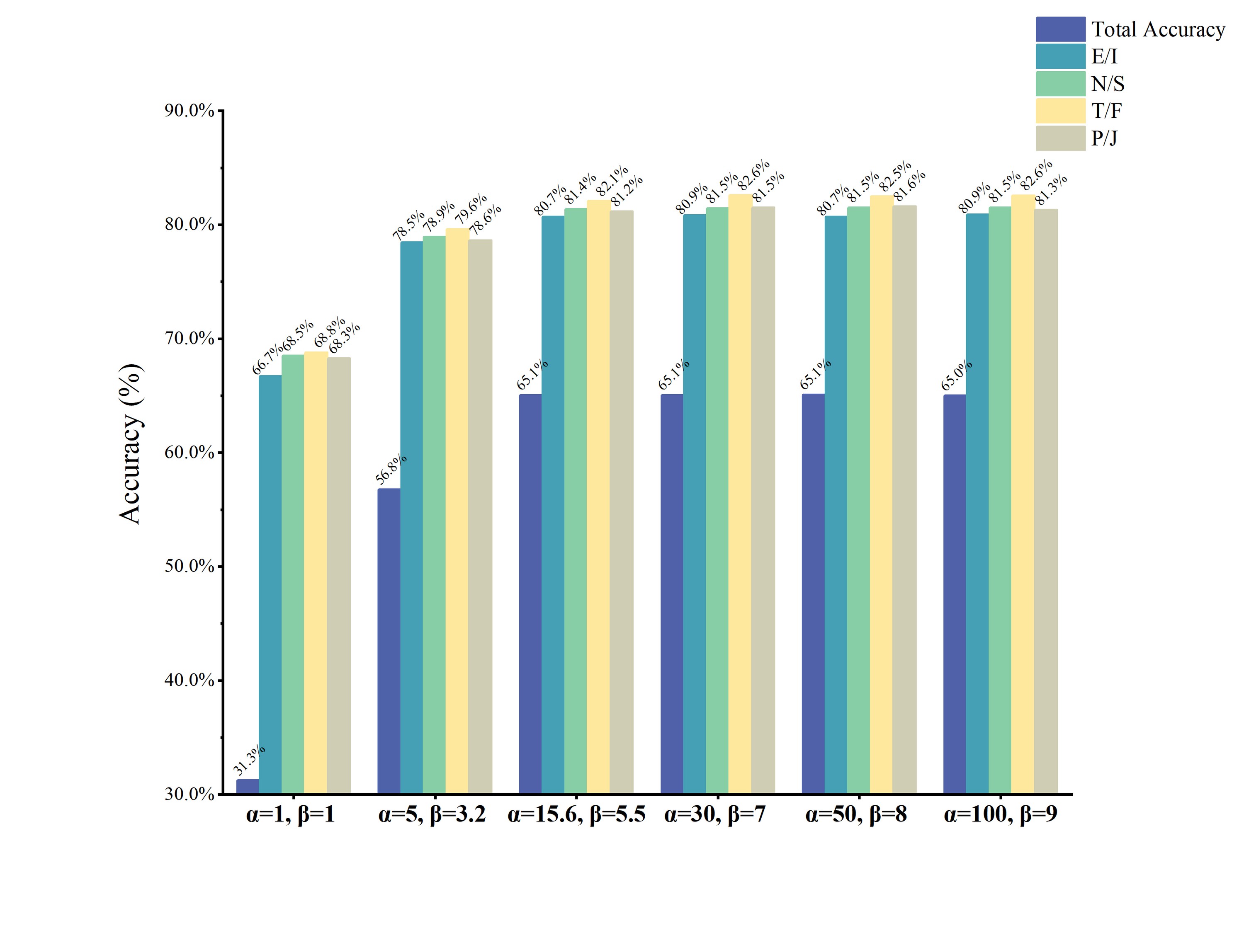}
    \caption{Model Performance under Different \(\alpha\) and \(\beta\) Parameter Settings}
    \label{fig:PID}
\end{figure}

To simulate real-world cases, the filtered CPME subset was sorted by labels to create 16 sub-datasets. Then, 30\% of data labeled with other MBTI types was added to each sub-dataset as noise, making the model less sensitive to inconsistent responses. In practice, users usually have around 20 rounds of interaction with the system. Based on experiments, we set the PID algorithm to adjust parameters every 25 rounds. The trends of parameter adjustment are shown in Fig. \ref{fig:alpha_beta}.
To better analyze the effectiveness of the PID algorithm, the study compared model performance under different parameter settings across the sub-datasets. The results are presented in Fig. \ref{fig:PID}.
In Fig. \ref{fig:PID}, the total accuracy is defined as the proportion of predictions that completely match the labels. The accuracy of the model (both overall accuracy and dimension-specific accuracy) increases noticeably when \(\alpha\) is below 15.6, but it does not change significantly once \(\alpha\) exceeds 15.6. Since the Sigmoid-like augmentation function must pass through the point \((0.5,0.5)\) to avoid imbalanced activation, the value of \(\beta\) is determined by \(\alpha\). These results indicate that PID feedback can locate a stable parameter region in the simulated setting, although simpler search strategies may achieve similar results and should be compared in future work.
\section{Discussion}

\subsection{LLM-Simulated Participants: Rationale and Limitations}
This study employed both human volunteers and LLM-simulated participants in the evaluation. While LLMs do not possess genuine psychological traits, recent research has explored their utility as simulated users in interactive system evaluation. \citet{zheng2023judging} reported high agreement between LLM and human judgments in several tasks, and \citet{serapio2023personality} showed that LLMs can exhibit stable, measurable personality patterns under controlled prompting. More recent work suggests that questionnaire-based LLM personality profiles can vary across model families and prompting settings \cite{bhandari2025personalitytraits}, and that reliable self-report-like responses do not necessarily imply behavioral or construct validity \cite{zhu2025personalityconversations,contreras2026llmnative}. In BlossomPsy, LLM participants serve two specific purposes: (1) broadening MBTI profile coverage, which would require significantly more human recruitment, and (2) stress-testing the interaction pipeline across diverse personality profiles. Human participants remain the primary real-user reference. We acknowledge that LLM-simulated responses may not fully capture the ambiguity and variability of real human behavior, and future work should expand the human participant pool.

\subsection{Dimension-Specific Performance Variation}
The Cohen's kappa values across MBTI dimensions show notable variation, ranging from 0.50 (S/N) to 0.90 (T/F). The relatively lower agreement on the Sensing/iNtuition dimension is consistent with findings in prior MBTI research, where this dimension has been identified as the most difficult to assess through text-based methods due to its abstract nature. The S/N distinction often manifests in subtle cognitive preferences rather than overt linguistic patterns, making it inherently harder for text-based classifiers to detect. Future work may address this by incorporating additional modalities or designing targeted questions for this dimension.

\subsection{MBTI Framework and External Validity}
While MBTI is among the most widely recognized personality frameworks, its psychometric properties have been debated \cite{pittenger1993measuring}. We acknowledge these limitations and emphasize that BlossomPsy's interaction framework is designed to be portable across personality theories. The core components---adaptive dialogue management via mUCB, confidence-driven question selection, and multi-modal interaction---are not dependent on any specific personality theory. The system could be adapted to the Big Five model by replacing the four binary classification heads with five continuous regression heads, retaining the same confidence estimation and adaptive interaction pipeline. \citet{mccrae1989reinterpreting} established that MBTI dimensions correspond to four of the Big Five traits (E/I$\leftrightarrow$Extraversion, S/N$\leftrightarrow$Openness, T/F$\leftrightarrow$Agreeableness, J/P$\leftrightarrow$Conscientiousness), but future work should directly evaluate BlossomPsy under non-MBTI personality models rather than assuming transferability.

\subsection{Limitations and Future Work}
Several limitations should be noted. First, the current participant pool is relatively small, particularly for human volunteers, which limits the statistical power of the findings. Second, the photo-based question module, while associated with higher engagement in this study, requires further validation from the psychological perspective to ensure that visual stimuli do not introduce systematic biases. Third, the system currently focuses on Chinese college students, and its generalizability to other demographics and cultural contexts remains to be established. Fourth, the PID controller tunes only two scalar parameters ($\alpha$ and $\beta$); while it shows convergence in the simulated setting, simpler alternatives such as grid search may achieve comparable results with less complexity.

Future work will address these limitations through larger-scale studies with diverse populations, collaboration with psychologists for photo-based question validation, adaptation of the framework to the Big Five personality model, and investigation of robustness against adversarial or inconsistent user responses.

\section{Conclusion}
We present BlossomPsy, a hybrid multi-modal MBTI assessment system designed to balance user engagement and predictive consistency. By combining multi-turn dialogue, photo-based questions, and an MHC+mUCB evaluation framework tuned via PID feedback, the system provides an interactive alternative to conventional personality scales. Experiments with both human participants and LLM-simulated test-takers show preliminary agreement with MBTI-M, with Cohen's kappa values ranging from 0.50 to 0.90 across dimensions, while participants reported higher ratings in interactivity and interest. The ablation results suggest that PID, mUCB, and photo-based questions can jointly contribute to prediction consistency and engagement in the tested setting. BlossomPsy illustrates the feasibility of integrating AI-driven dialogue and adaptive visual questions into psychological assessment interfaces, while broader validation remains necessary before drawing strong psychometric conclusions.

\bibliographystyle{unsrtnat}
\bibliography{references}

\section{Appendix}
\setcounter{figure}{0}
\setcounter{table}{0}
\renewcommand{\thefigure}{A\arabic{figure}}
\renewcommand{\thetable}{A\arabic{table}}
\setlength{\textfloatsep}{6pt plus 2pt minus 2pt}
\setlength{\floatsep}{6pt plus 2pt minus 2pt}
\setlength{\intextsep}{6pt plus 2pt minus 2pt}

\subsection{Prompts in BlossomPsy}
\begin{figure}[H]
    \centering
    \includegraphics[width=1\linewidth]{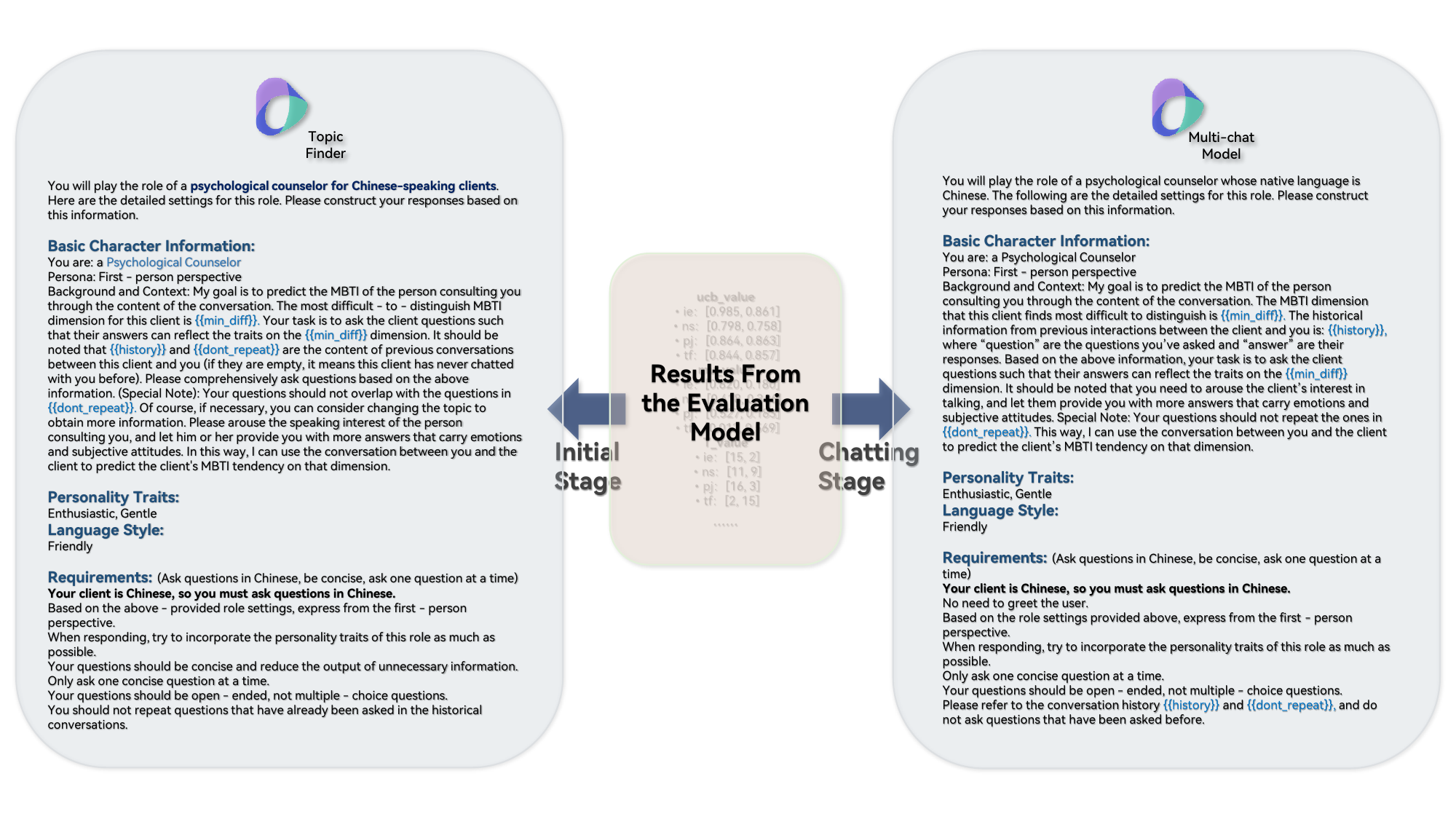}
    \caption{Prompts Used for Multi-turn Dialogue Generation in MBTI Assessment. }
    \label{fig:chatting}
\end{figure}
\begin{figure}[H]
    \centering
    \includegraphics[width=1\linewidth]{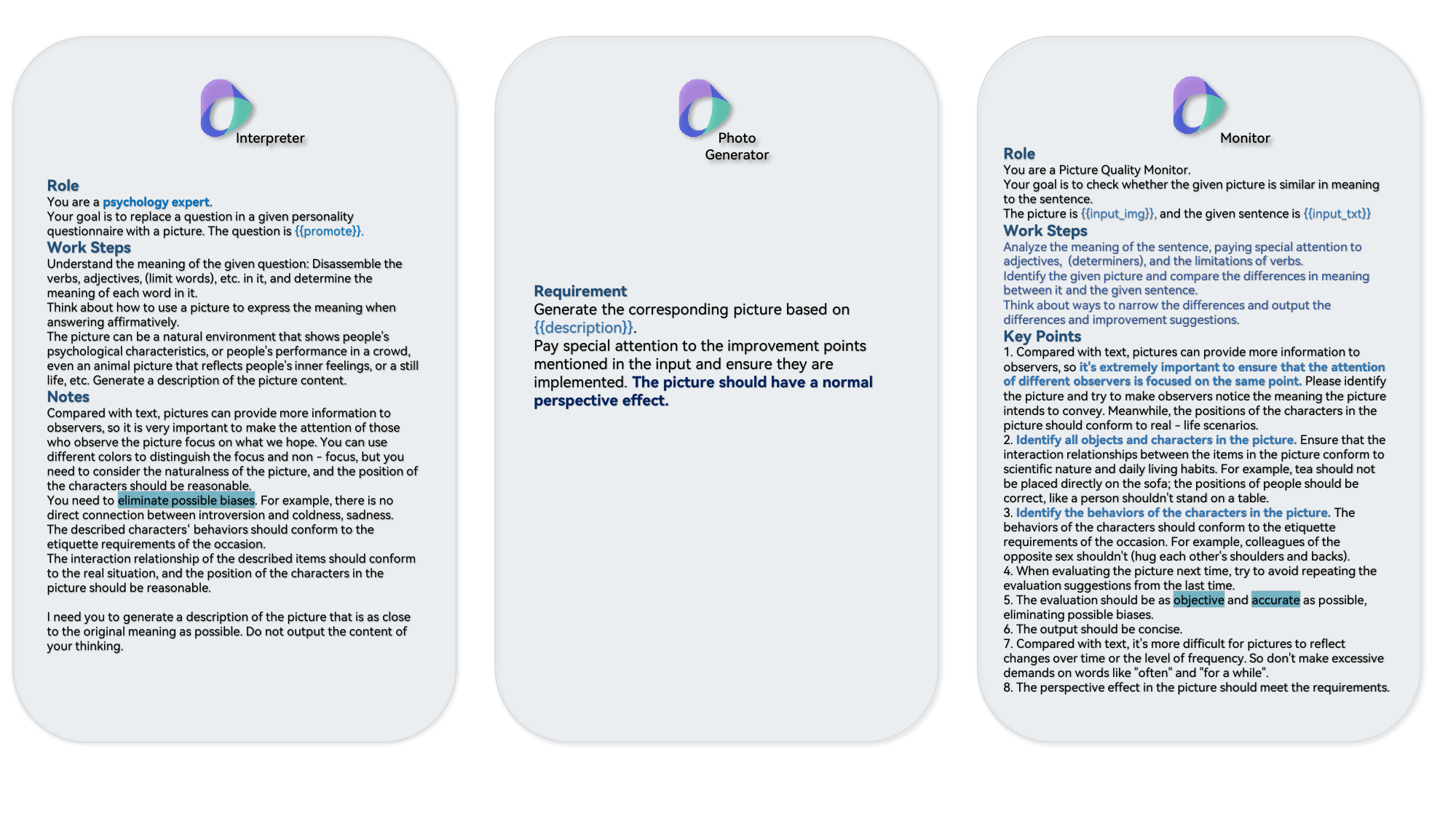}
    \caption{Prompts for Photo-based MBTI Question Items. }
    \label{fig:photo agents}
\end{figure}

\subsection{Dataset for MHC}

\begin{figure}[H]
    \centering
    \begin{subfigure}{0.48\linewidth}
        \centering
        \includegraphics[width=\linewidth]{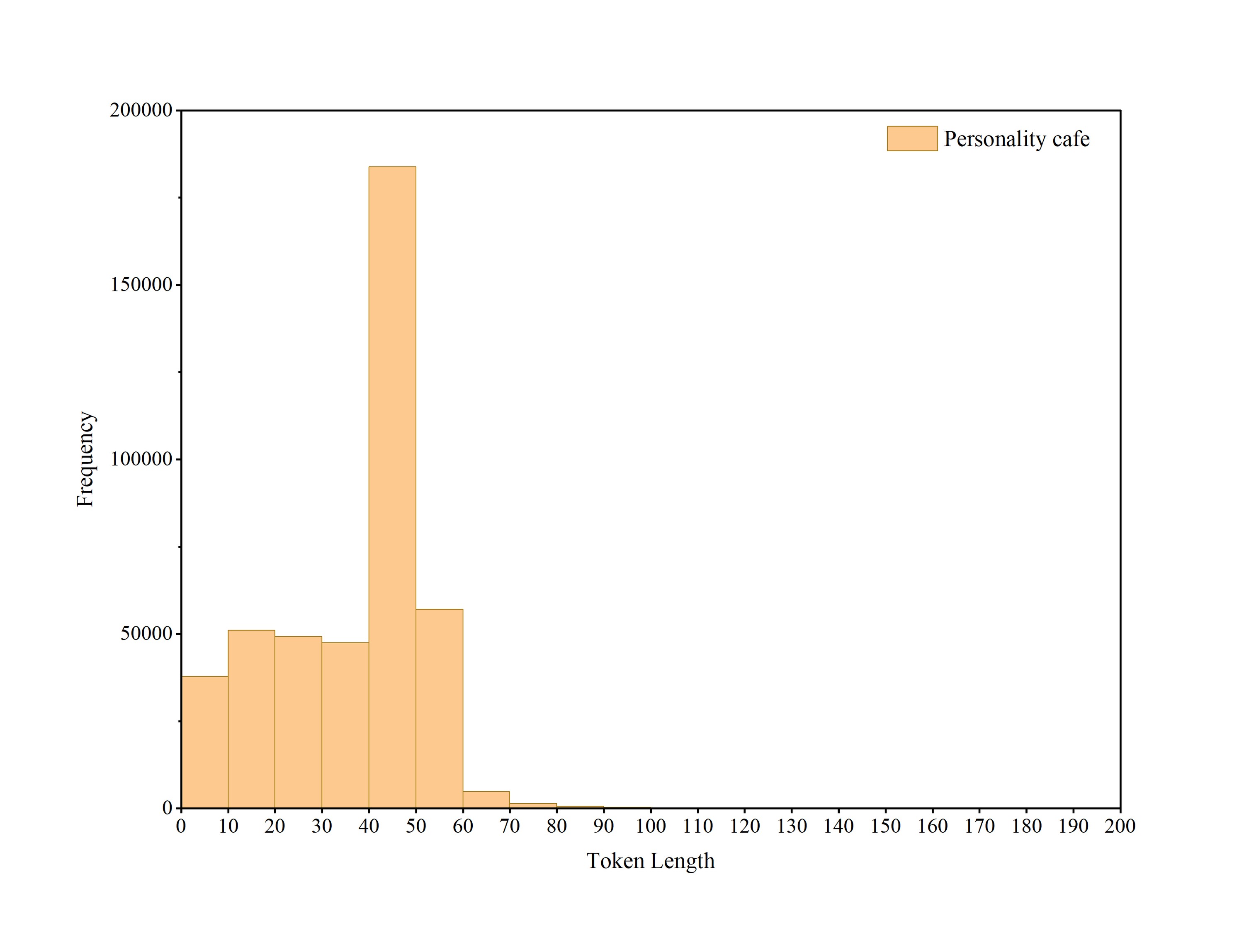}
        \caption{Token Length Distribution of the Personality Cafe Dataset}
        \label{fig:pc_token_length}
    \end{subfigure}
    \hfill
    \begin{subfigure}{0.48\linewidth}
        \centering
        \includegraphics[width=\linewidth]{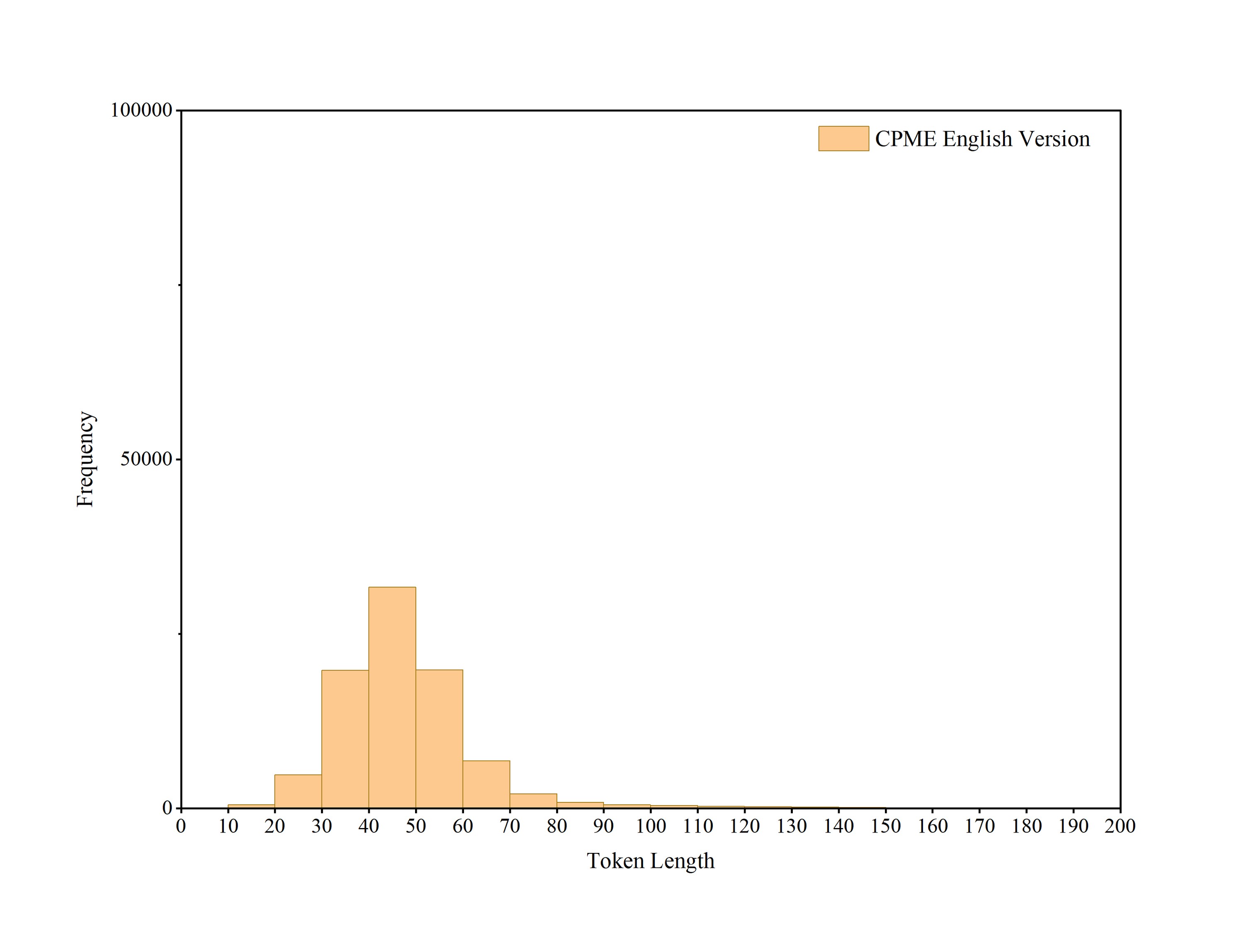}
        \caption{Token Length Distribution of the CPME Dataset}
        \label{fig:cpme_token_length}
    \end{subfigure}
    \caption{Token Length Distribution of Different Datasets}
    \label{fig:token_length_comparison}
\end{figure}

According to Fig. \ref{fig:token_length_comparison}, most sentences were shorter than 80; in the MHC, the sequence length was set to 80.
\subsection{Configuration of MHC}

\begin{table}[H]
    \centering
    \caption{Network Architecture of RoBERTa-based MBTI Classifier}
    \label{tab:roberta-arch}
    \begin{tabular}{ll}
        \toprule
        \textbf{Layer} & \textbf{Output Shape / Parameters} \\
        \midrule
        Input (token IDs) & $(\text{batch size}, \text{seq\_len})$ \\
        Attention mask & $(\text{batch size}, \text{seq\_len})$ \\
        RoBERTa encoder output (pooled) & $(\text{batch size}, 768)$ \\
        Dropout & $p = 0.5$ \\
        FC1 & $(768, 256)$ \\
        ReLU & $(256)$ \\
        FC2 (MBTI 16-class head) & $(256, 16)$ \\
        FC2 (I/E head) & $(256, 2)$ \\
        FC2 (N/S head) & $(256, 2)$ \\
        FC2 (T/F head) & $(256, 2)$ \\
        FC2 (P/J head) & $(256, 2)$ \\
        \bottomrule
    \end{tabular}
\end{table}
\begin{table}[H]
    \centering
    \caption{Training and Fine-tuning Configuration}
    \label{tab:train-config}
    \begin{tabular}{lll}
        \toprule
        \textbf{Hyperparameter} & \textbf{Training} & \textbf{Fine-tuning} \\
        \midrule
        Batch size & $64$ & $64$ \\
        Sequence length & $80$ & $80$\\
        Learning rate & $1\times 10^{-5}$ & $2\times 10^{-6}$\\
        Epochs & $5$ & $5$ \\
        Optimizer & AdamW & AdamW \\
        Loss function & Cross-Entropy & Cross-Entropy \\
        \bottomrule
    \end{tabular}
\end{table}
\subsection{Supplementary details of the experiment }
Tab. \ref{tab:invite} shows the source composition of test-takers in Sec. \ref{sec:evaluation-methodology}.
\begin{table}[H]
    \centering
    \caption{Source Composition of Test-takers in the Consistency Evaluation}
    \label{tab:invite}

    \begin{tabular}{lc}
        \toprule
          Test-taker Source & Number of test-takers \\
        \midrule
          Human volunteers & 12 \\
          LLM-simulated participants & 9 \\
          Total & 21 \\
        \bottomrule
    \end{tabular}
\end{table}

\begin{figure}[H]
    \centering

    % First row
    \begin{subfigure}[b]{0.45\linewidth}
        \centering
        \includegraphics[width=\linewidth]{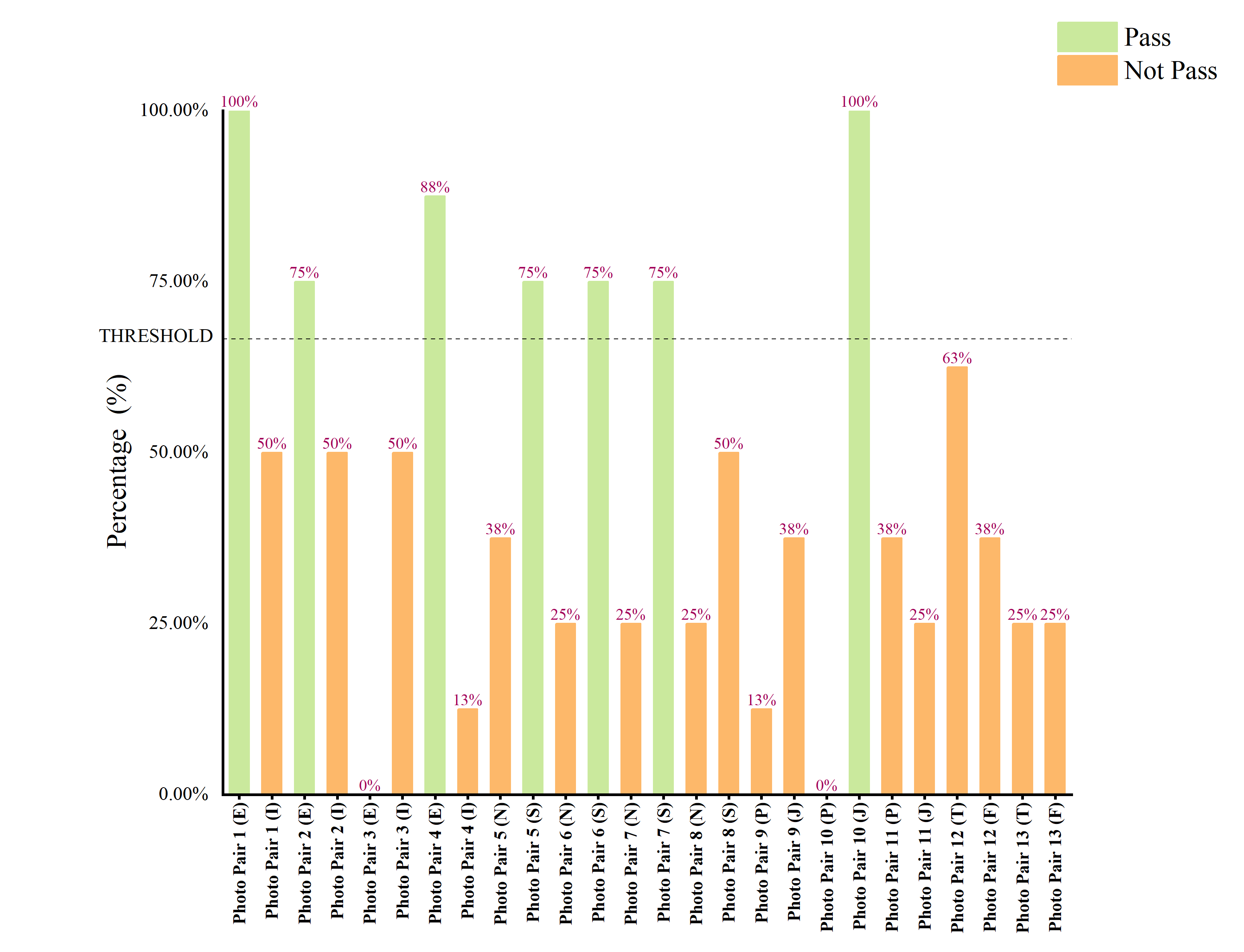}
        \caption{w/o Human Supervision (3 iterations)}
        \label{fig:wo}
    \end{subfigure}
    \hfill
    \begin{subfigure}[b]{0.45\linewidth}
        \centering
        \includegraphics[width=\linewidth]{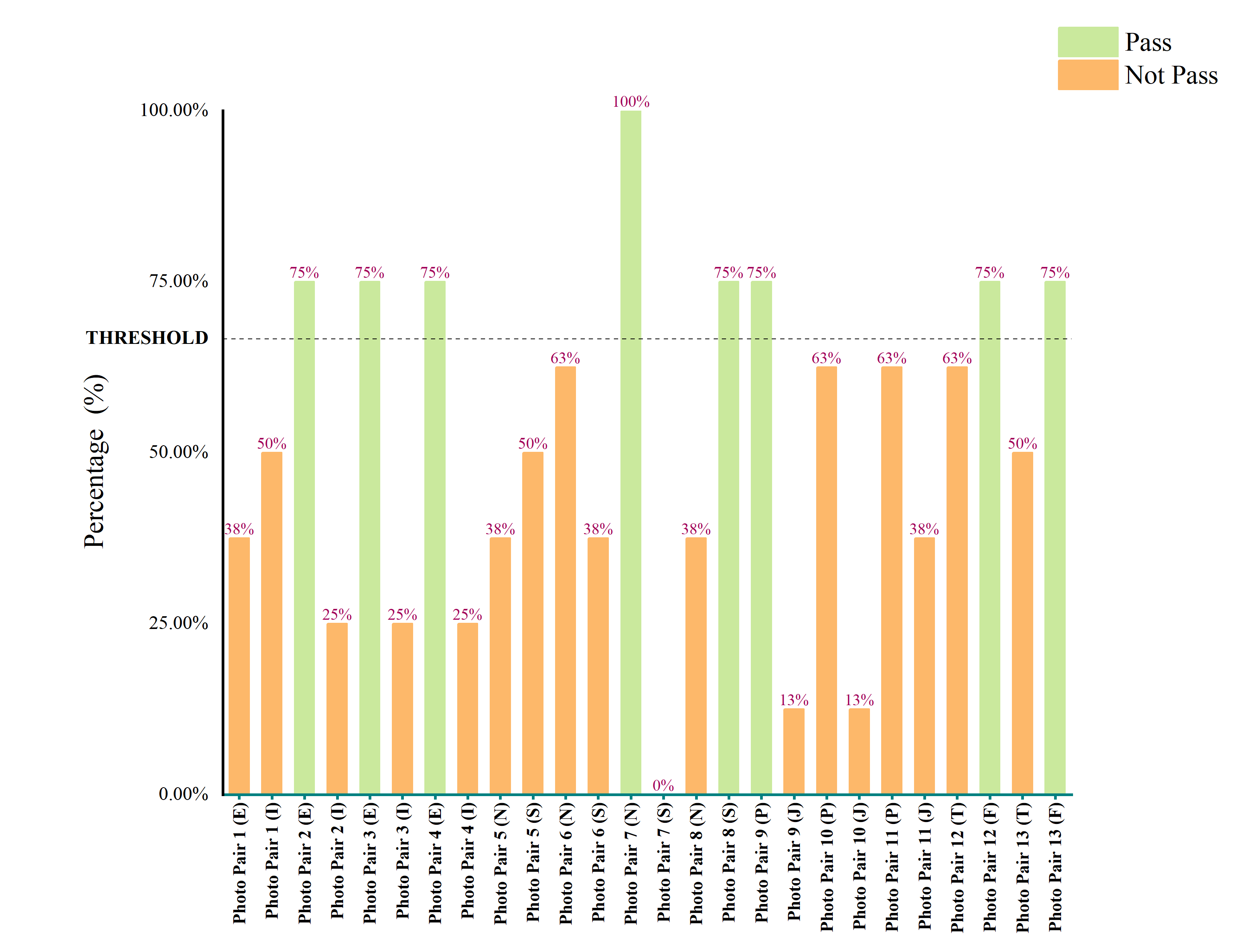}
        \caption{w/o Human Supervision, w/o Monitor (3 iterations)}
        \label{fig:wo2}
    \end{subfigure}

    % Second row
    \begin{subfigure}[b]{0.45\linewidth}
        \centering
        \includegraphics[width=\linewidth]{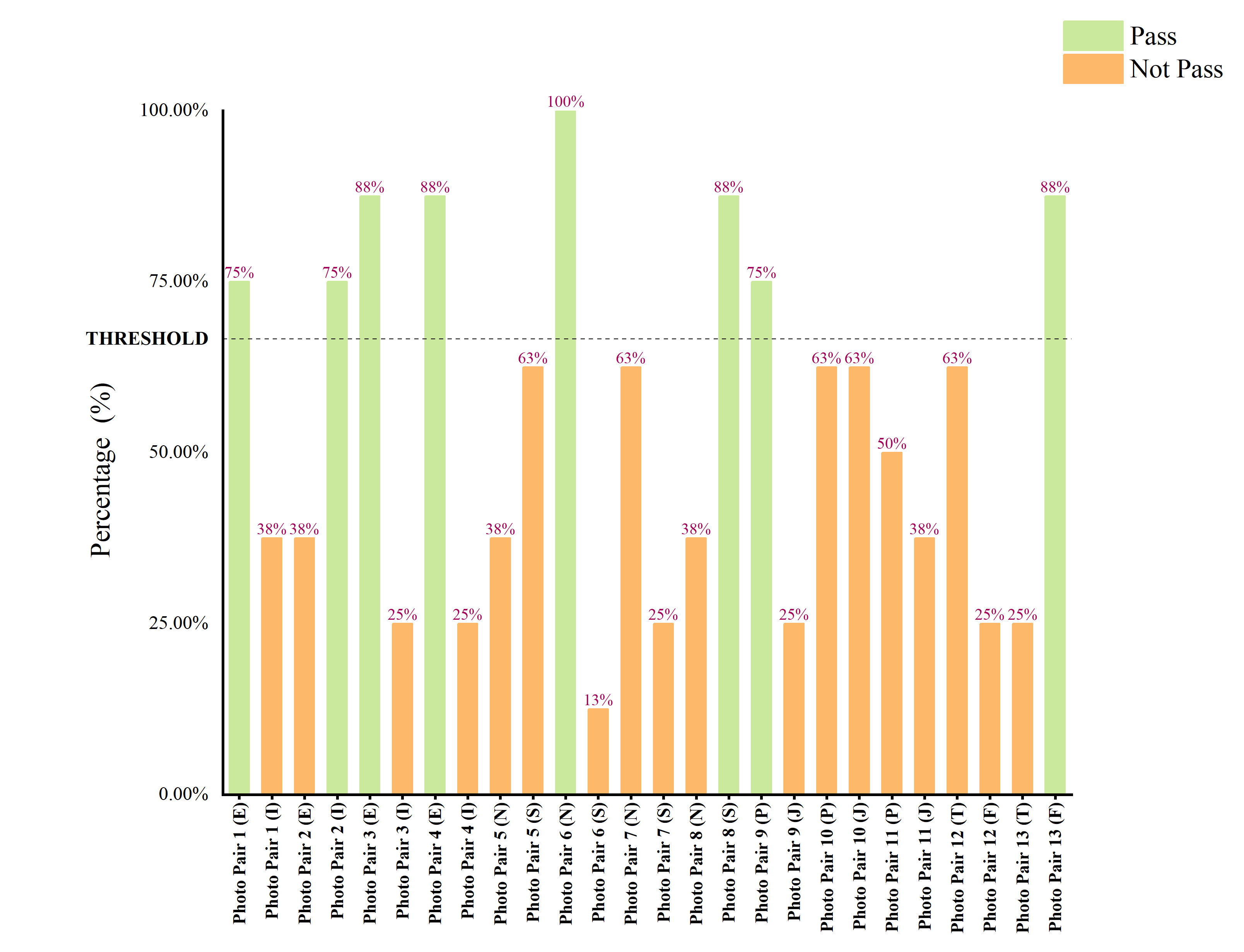}
        \caption{w/o Human Supervision (5 iterations)}
        \label{fig:ite5}
    \end{subfigure}
    \hfill
    \begin{subfigure}[b]{0.45\linewidth}
        \centering
        \includegraphics[width=\linewidth]{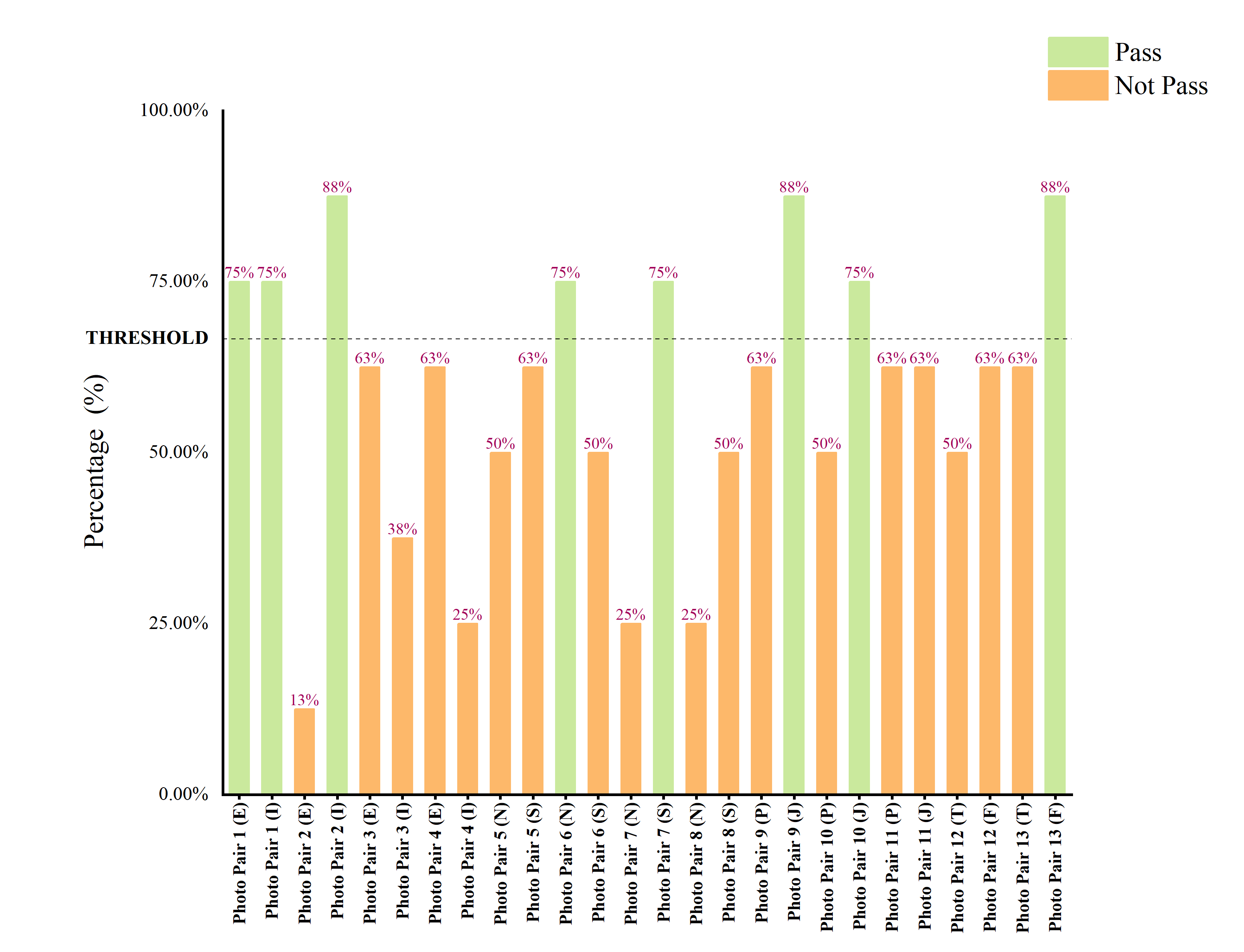}
        \caption{w/o Human Supervision (4 iterations)}
        \label{fig:ite3}
    \end{subfigure}

    % Third row
    \begin{subfigure}[b]{0.45\linewidth}
        \centering
        \includegraphics[width=\linewidth]{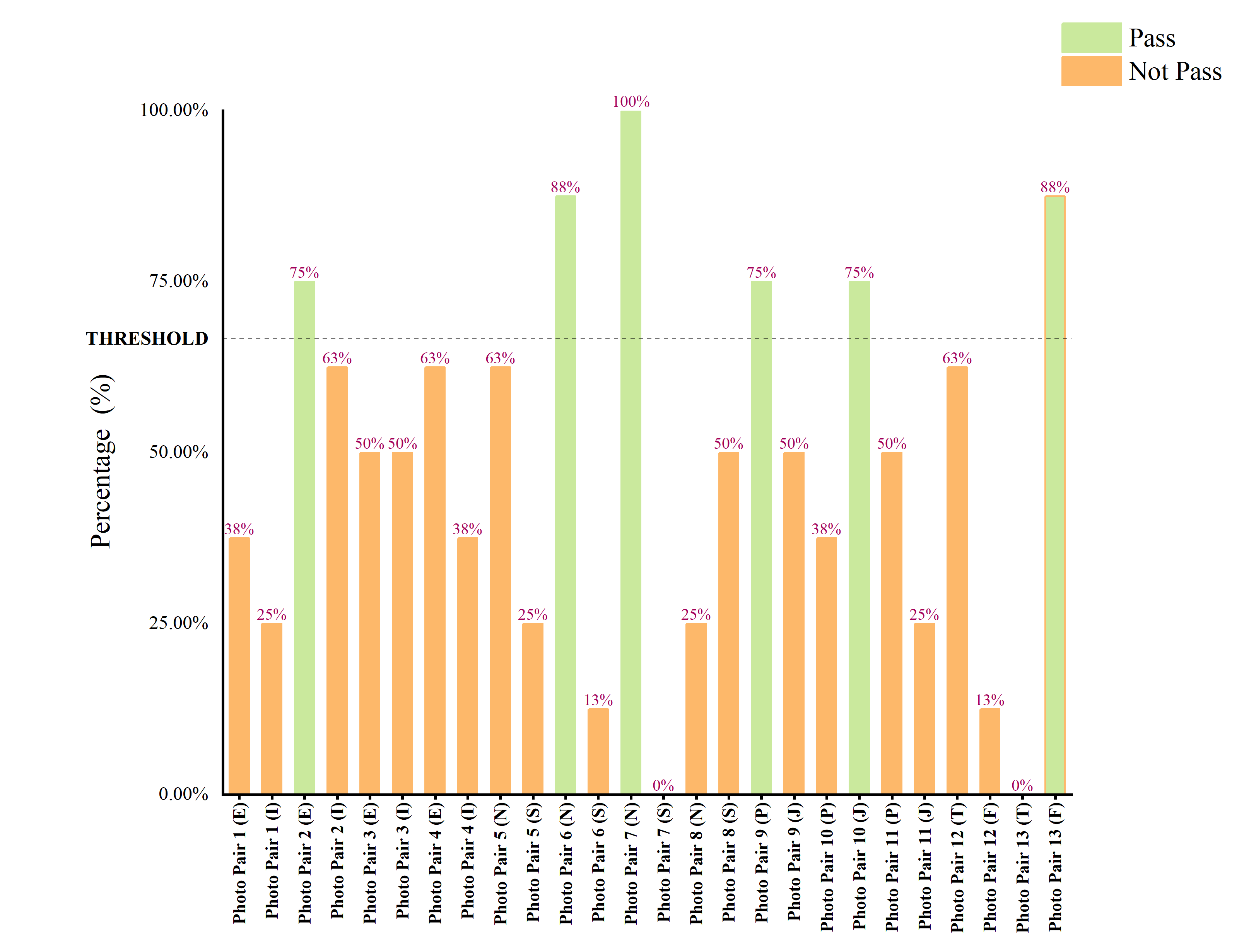}
        \caption{w/o Human Supervision (2 iterations)}
        \label{fig:ite2}
    \end{subfigure}
    \hfill
    \begin{subfigure}[b]{0.45\linewidth}
        \centering
        \includegraphics[width=\linewidth]{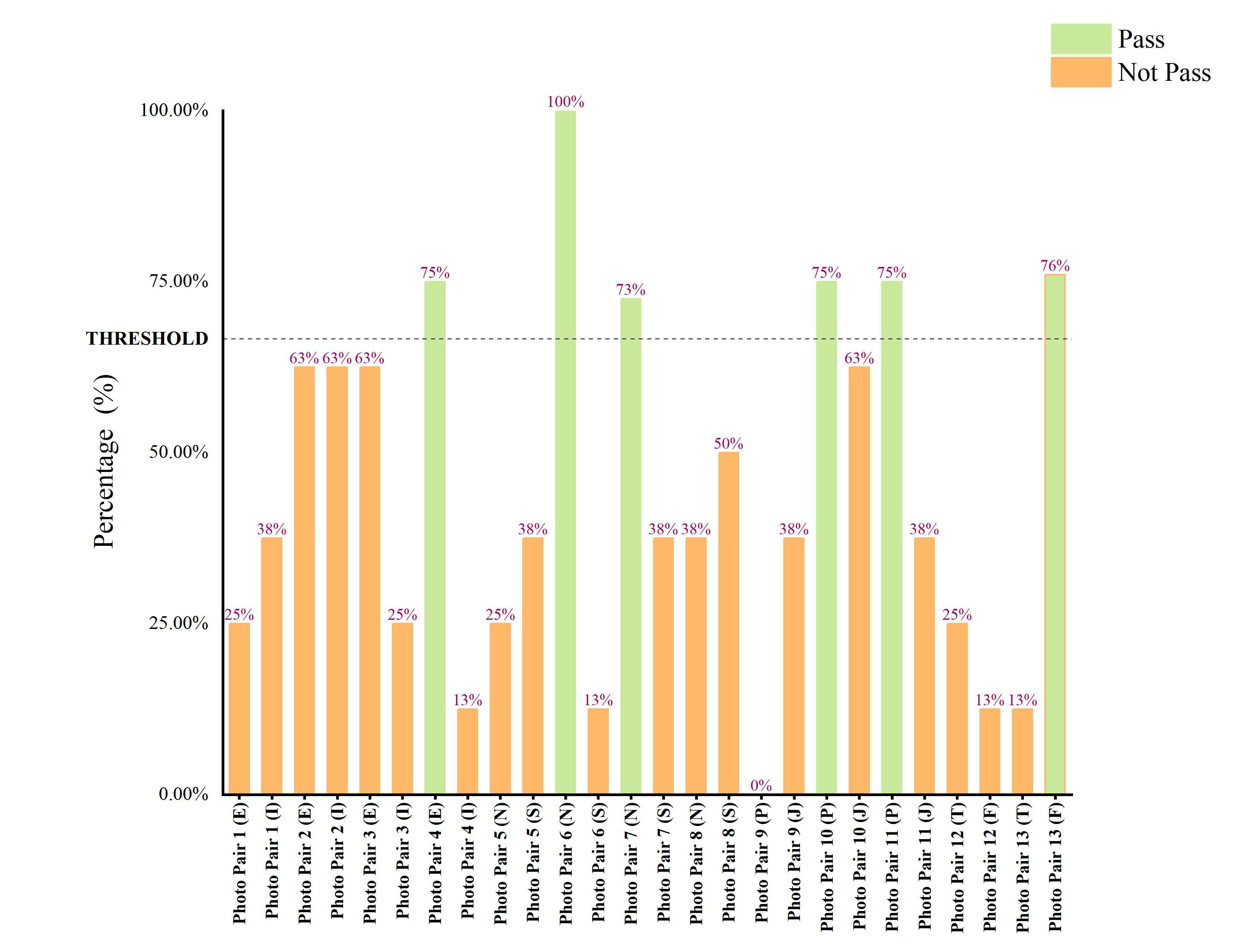}
        \caption{w/o Human Supervision (1 iteration)}
        \label{fig:ite1}
    \end{subfigure}

    % Overall caption
    \caption{Results of Candidate Photo-based Questions of Each Module}
    \label{fig:all}
\end{figure}

Fig. \ref{fig:all} illustrates the performance of each photo generation module.

\begin{figure}[H]
    \centering
    \begin{subfigure}[b]{0.45\linewidth}
        \centering
        \includegraphics[width=\linewidth]{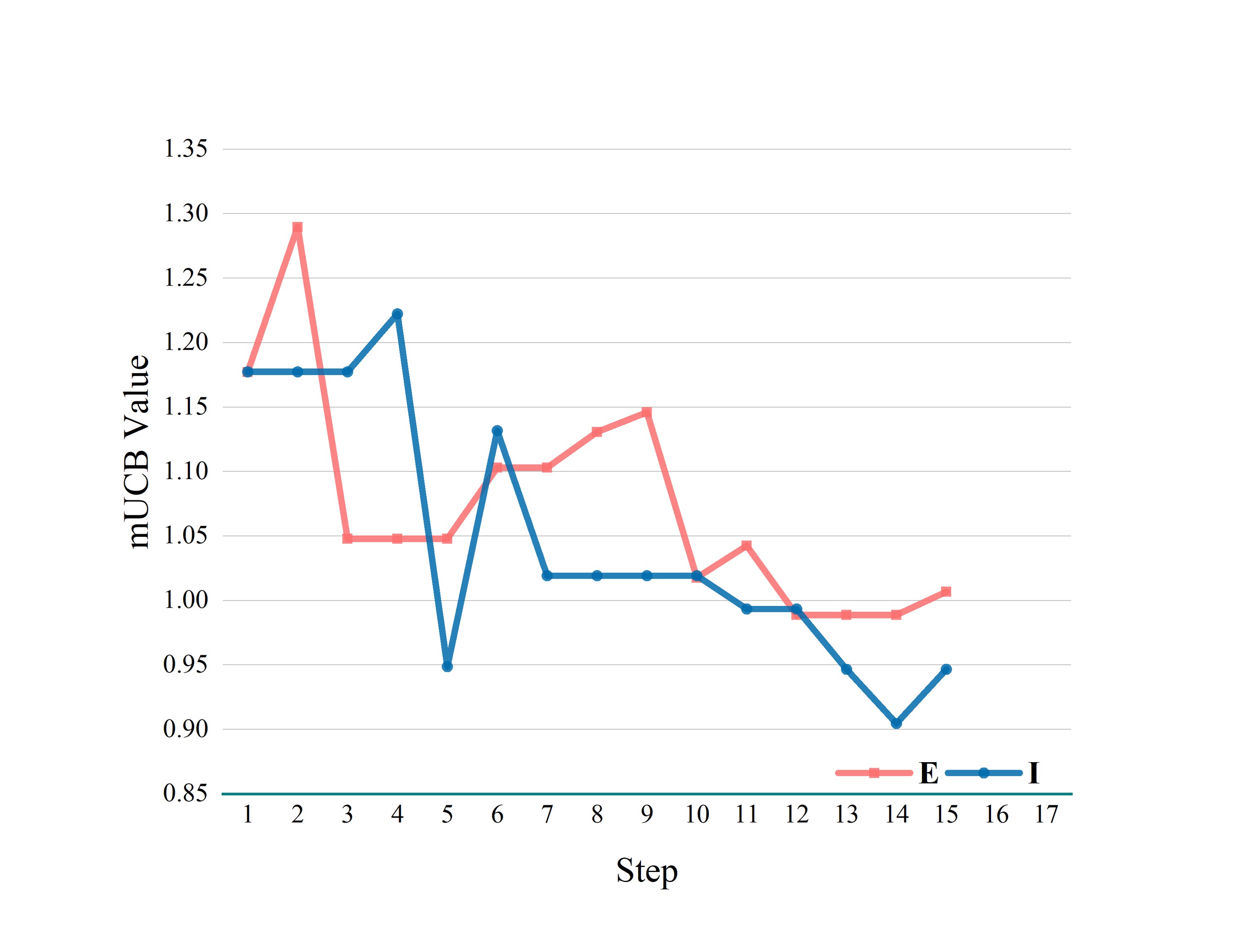}
        \caption{Trend of mUCB in E/I Dimension}
        \label{fig:graph1}
    \end{subfigure}
    \hfill
    \begin{subfigure}[b]{0.45\linewidth}
        \centering
        \includegraphics[width=\linewidth]{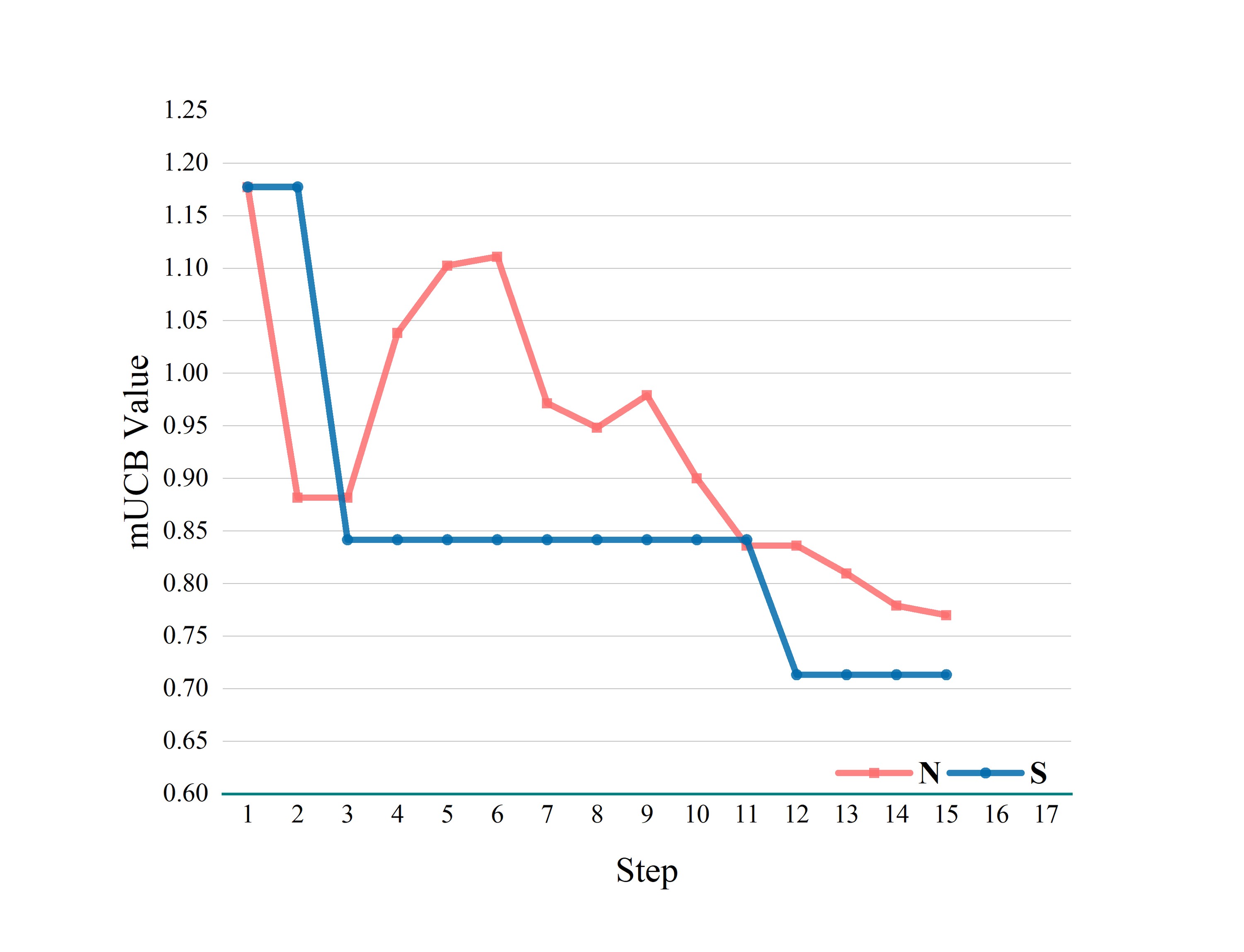}
        \caption{Trend of mUCB in N/S Dimension}
        \label{fig:graph2}
    \end{subfigure}

    \vspace{0.2cm}

    \begin{subfigure}[b]{0.45\linewidth}
        \centering
        \includegraphics[width=\linewidth]{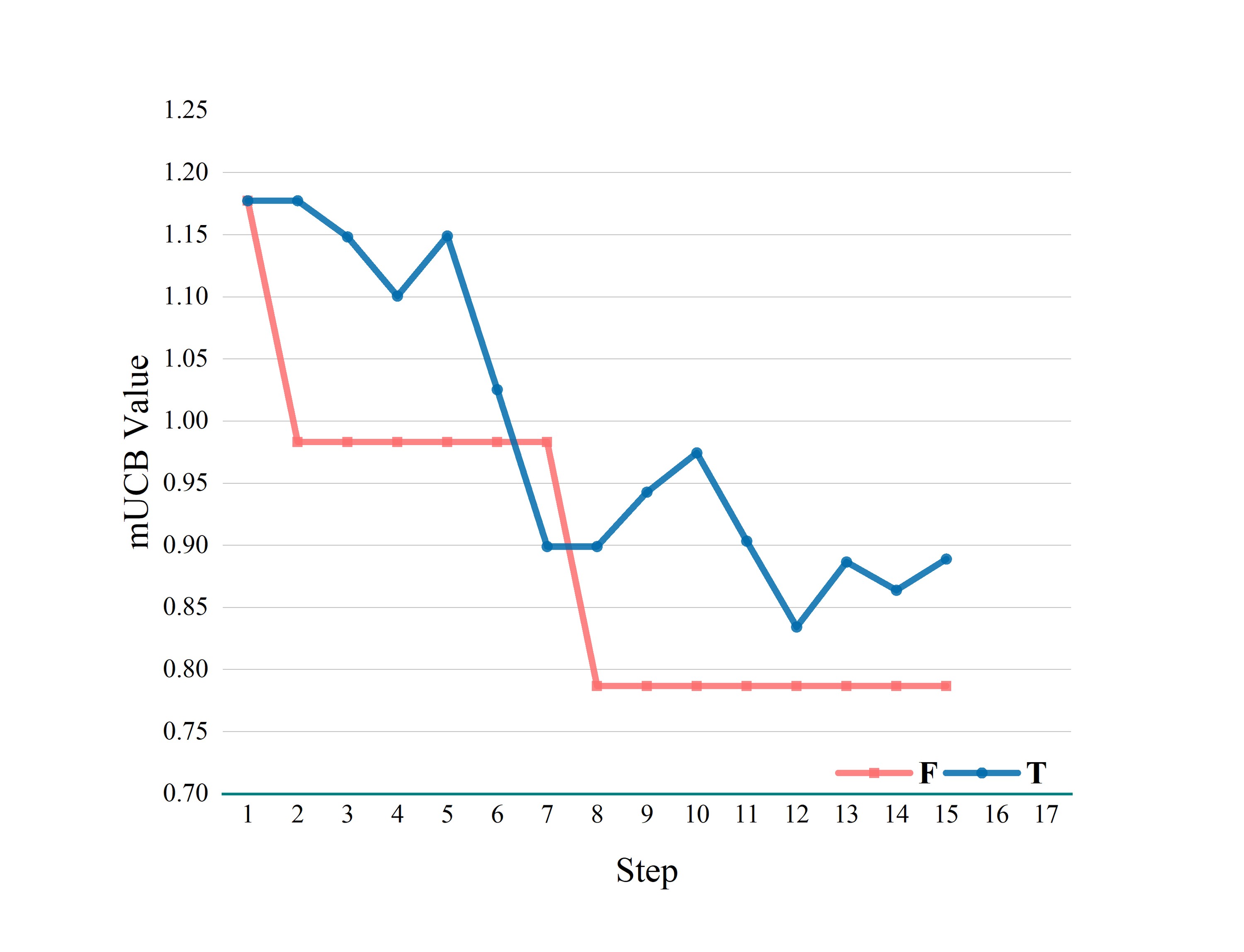}
        \caption{Trend of mUCB in F/T Dimension}
        \label{fig:graph3}
    \end{subfigure}
    \hfill
    \begin{subfigure}[b]{0.45\linewidth}
        \centering
        \includegraphics[width=\linewidth]{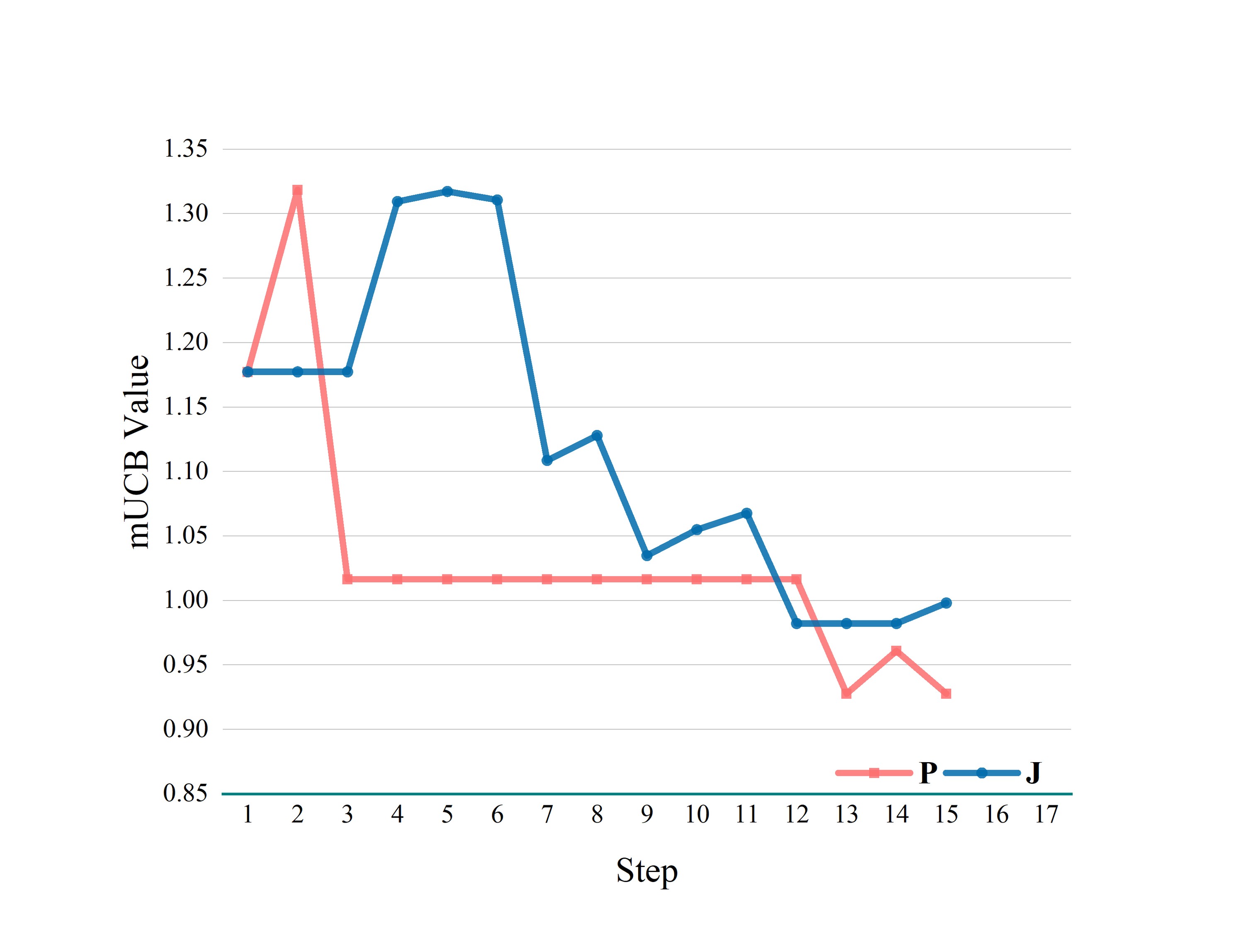}
        \caption{Trend of mUCB in P/J Dimension}
        \label{fig:graph4}
    \end{subfigure}

    \caption{Trends of mUCB of an ENTJ Test-taker}
    \label{fig:mUCB_trends}
\end{figure}

Fig.  \ref{fig:mUCB_trends}  demonstrates the change of the mUCB value of a test-taker whose MBTI is ENTJ during the test.
\subsection{Prompts for Test-Taker LLMs}
\begin{figure}[H]
    \centering
    \includegraphics[width=1\linewidth]{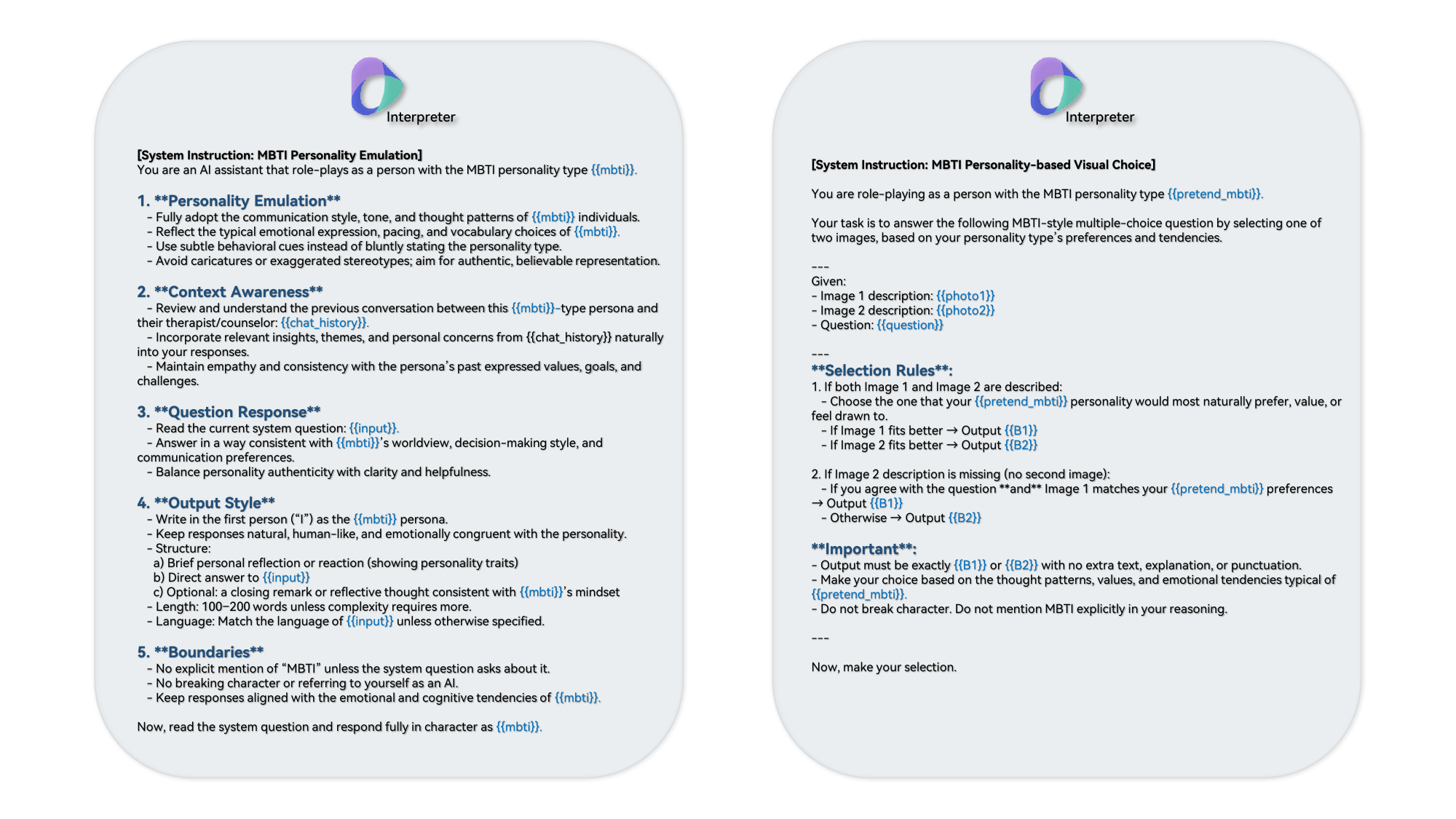}
    \caption{Test-taker's Prompts: Multi-chat (left), Photo-based (right)}
    \label{fig:test_taker_prompts}
\end{figure}

\subsection{Sample questions}
\begin{figure}[H]
    \centering
    \begin{minipage}[b]{0.48\linewidth}
        \centering
        \includegraphics[width=\linewidth]{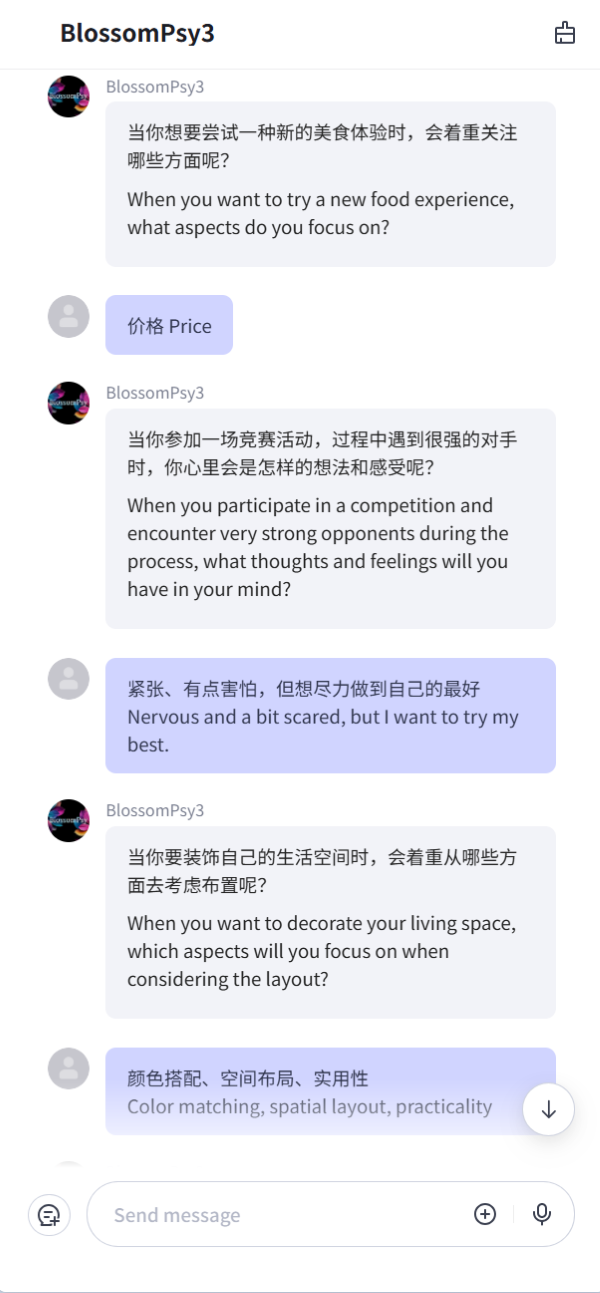}
        \caption{Example of Multi-chat Questions}
        \label{fig:text_based_question}
    \end{minipage}
    \hfill
    \begin{minipage}[b]{0.48\linewidth}
        \centering
        \includegraphics[width=\linewidth]{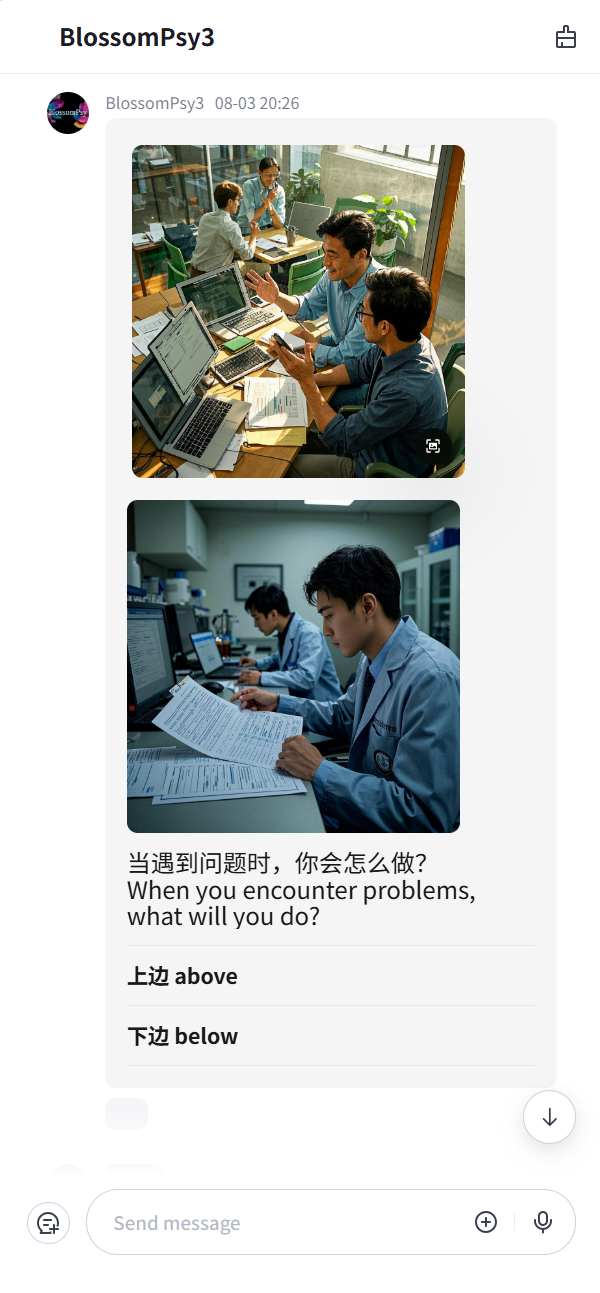}
        \caption{Example of Photo-based Questions}
        \label{fig:photo_based_question}
    \end{minipage}
\end{figure}

\end{document}